\documentclass[conference]{IEEEtran}
\IEEEoverridecommandlockouts
\usepackage[utf8]{inputenc}
\usepackage{cite}
\usepackage{amsmath,amssymb,amsfonts}
\usepackage{mathrsfs}
\usepackage{graphicx}
\usepackage{textcomp}
\usepackage{xcolor}
\usepackage{bm}
\usepackage{mathtools}
\usepackage[noend]{algpseudocode}
\usepackage{algorithmicx,algorithm}
\usepackage{subfigure}
\usepackage{tikz}

\newtheorem{thm}{Theorem}

\newtheorem{prop}{Proposition}
\newtheorem{lem}{Lemma}

\newtheorem{rmk}{Remark}

\def\BibTeX{{\rm B\kern-.05em{\sc i\kern-.025em b}\kern-.08em
		T\kern-.1667em\lower.7ex\hbox{E}\kern-.125emX}}
\begin{document}
\title{Fairness-aware Regression Robust to Adversarial Attacks}
\author{\IEEEauthorblockN{ Yulu Jin and Lifeng Lai\thanks{Y. Jin and L. Lai are with the Department of Electrical and Computer Engineering, University of California, Davis, CA. Email: 	\{yuljin,lflai\}@ucdavis.edu. The work of Y. Jin and L. Lai was supported by the National Science Foundation under Grants CNS-1824553, CCF-1908258 and ECCS-2000415. This paper has been submitted in part to 2023 IEEE International Conference on Acoustics, Speech, and Signal Processing (ICASSP) \cite{jin2023fair}.}
}}
\maketitle
	\begin{abstract}
		In this paper, we take a first step towards answering the question of how to design fair machine learning algorithms that are robust to adversarial attacks. Using a minimax framework, we aim to design an adversarially robust fair regression model that achieves optimal performance in the presence of an attacker who is able to add a carefully designed adversarial data point to the dataset or perform a rank-one attack on the dataset. By solving the proposed nonsmooth nonconvex-nonconcave minimax problem, the optimal adversary as well as the robust fairness-aware regression model are obtained. For both synthetic data and real-world datasets, numerical results illustrate that the proposed adversarially robust fair models have better performance on poisoned datasets than other fair machine learning mod{\tiny }els in both prediction accuracy and group-based fairness measure. 
	\end{abstract}

\section{Introduction} \label{sec:intro}
Machine learning models have been used in various domains, including several security and safety critical applications, such as banking, education, healthcare, law enforcement etc. However, it has been shown that machine learning algorithms can mirror or even amplify biases against population subgroups \cite{corbett2018measure,martinez2020minimax}, for example, based on race or sex. With direct social and economic impact on individuals, it is imperative to build ML models ethically and responsibly to avoid these biases. To this end various algorithms have been developed to find fair machine models (FML) that satisfy different fairness measures \cite{goel2018non,zhao2019inherent,agarwal2019fair,chzhen2020fair,9496158,roh2020fr,chi2021understanding}. 

In the meantime, a large body of work has shown that machine learning models are vulnerable to various types of attacks \cite{chen2017targeted,shafahi2018poison,9298957,9551762}. Thus, a major and natural concern for fair machine learning algorithms is their robustness in adversarial environments. Recent works show that well-designed adversarial samples can significantly reduce the test accuracy as well as exacerbating the fairness gap of ML models~\cite{solans2020poisoning,mehrabi2020exacerbating,van2021poisoning,chang2020adversarial}. 

In light of the vulnerabilities of existing fair machine learning algorithms, there is a pressing need to design fairness-aware learning algorithms that are robust to adversarial attacks. As the first step towards this goal, we focus on regression problems and design a fair regression model that is robust to adversarial attacks. In particular, we consider two increasingly complex attack models. We first consider a scenario where the adversary is able to add one carefully designed adversarial data point to the dataset. We then consider a more powerful adversary who can directly modify the existing data points in the feature matrix. Particularly, we consider a rank-one modification attack, where the attacker carefully designs a rank-one matrix and adds it to the existing data matrix. 

To design the robust fairness-aware model, we formulate a game between a defender aiming to minimize the accuracy loss and bias, and an attacker aiming to maximize these objectives. To characterize both the prediction and fairness performance of a model, the objective function is selected to be a combination of prediction accuracy loss and group fairness gap. Since the goals of the adversary and the fairness-aware defender are opposite, a minimax framework is introduced to characterize the considered problem. By solving the minimax problem, the optimal adversary as well as the robust fair regression model can be derived. 

To solve the problem, one major challenge is that the proposed minimax problem is nonsmooth nonconvex-nonconcave, which may not have a local saddle point in general \cite{jiang2022optimality}. Although there exist many iterative methods for finding stationary points or local optima of nonconvex-concave or nonconvex-nonconcave minimax problems \cite{lin2020gradient,yang2020global,9070155,diakonikolas2021efficient,mangoubi2021greedy,lee2021fast,ostrovskii2021nonconvex}, there are usually specific assumptions that are not satisfied in our proposed realistic problems. To solve the complicated minimax problems in hand, we carefully examine the underlying structure of the inner maximization problem and the outer minimization problem, and then exploit the identified structure to design efficient algorithms. 

For the scenario where the adversary adds a poisoned data point into the dataset, when solving the inner maximization problem, we deal with the non-smooth nature of the objective function and obtain a structure that characterizes the best adversary, which is a function of the regression coefficient $\bm{\beta}$ of the defense model. We then analyze the minimization problem by transforming it to four sub-problems where each sub-problem is a non-convex quadratic minimization problem with multiple quadratic constraints, which is usually NP hard \cite{luo2007approximation,huang2016consensus}, and finding a global minimizer is very challenging. By exploring the underlying properties of a specific sub-problem, we investigate $8$ different cases, and obtain a global minimizer to such sub-problem. Then the minimum point of the proposed four sub-problems, $\bm{\beta}^*_{rob}$, corresponds to the optimal robust fairness-aware model, and the best adversarial data sample is obtained by fitting $\bm{\beta}^*_{rob}$ to the derived optimal attack strategy. On both synthetic data and real-world datasets, numerical results illustrate that the proposed robust fairness-aware regression model has better performance than the unrobust fair model as well as the ordinary linear regression model in both prediction accuracy and group-based fairness.

For the rank-one attack scheme, we transform the maximization problem into a form with five arguments, four of which can be solved exactly. With this transformation, the original nonconvex-nonconcave minimax problem for two vectors can be converted into several weakly-convex-weakly-concave minimax problems for one vector and one scalar, which can be approximately solved using existing algorithms such as \cite{liu2021first}. With the proposed algorithm, the optimal attack scheme of the adversary and the adversarially robust fairness-aware model can be obtained simultaneously. On two real-world datasets, numerical results illustrate that the performance of the adversarially robust model relies on the trade-off parameter between prediction accuracy and fairness guarantee. By properly choosing such parameter, the robust model can achieve desirable performance in both prediction accuracy and group-based fairness. On the other hand, for other fair regression models, at least one performance metric will be severely affected by the rank-one attack.

This journal paper is an extension of conference paper \cite{jin2023fair}. In addition to the rank-one attack considered in~\cite{jin2023fair}, in this journal paper, we also explore another attack scheme where additional data samples can be added to the existing dataset. We carefully design the feature vector, outcome variable, and group membership index of the poisoned sample to explore the impact of such attack and derive the robust fairness-aware model. In addition, we conduct more comprehensive numerical simulations and provide detailed theoretical analysis and proofs.  

The remainder of the paper is organized as follows. In Section~\ref{sec:related}, we summarize the related work of this paper. In Section~\ref{sec:one point}, we investigate the case when the adversary is allowed to add a poisoned data point into the dataset. In Section~\ref{sec:rank-one}, we consider a more powerful adversary who is able to perform a rank-one attack on the dataset. In Section~\ref{sec:numerical}, we present numerical results. Finally, we offer concluding remarks in Section~\ref{sec:conclude}.

\section{Related work} \label{sec:related}
\noindent\textbf{Adversarial attacks on FML.} There are many research works exploring the design of adversarial examples to reduce the testing accuracy and fairness of FML models. For example, \cite{solans2020poisoning} develops a gradient-based poisoning attack, \cite{mehrabi2020exacerbating} presents anchoring attack and influence attack, \cite{van2021poisoning} provides three online attacks based on different group-based fairness measures, and \cite{chang2020adversarial} shows that adversarial attacks can worsen the model’s fairness gap on test data while satisfying the fairness constraint on training data. \\
\textbf{Adversarial robustness.} A large variety of methods have been proposed to improve the model robustness against adversarial attacks \cite{kurakin2016adversarial,madry2017towards,zhang2019theoretically,9024000}. Although promising to improve the model’s robustness, those adversarial training algorithms have been observed to result in a large disparity of accuracy and robustness among different classes while natural training does not present a similar issue \cite{xu2021robust}. \\
\textbf{Intersection of fairness and robustness.} Fairness and robustness are critical elements of trustworthy AI that need to be addressed together \cite{roh2021sample}. Firstly, in the field of adversarial training, several research works are proposed to interpret the accuracy/robustness disparity phenomenon and to mitigate the fairness issue \cite{wadsworth2018achieving,madras2018learning,roh2021sample}. For example, \cite{wadsworth2018achieving} presents an adversarially-trained neural network that is closer to achieve some fairness measures than the standard model on the Correctional Offender Management Profiling for Alternative Sanctions (COMPAS) dataset. Secondly, a class-wise loss re-weighting method is shown to obtain more fair standard and robust classifiers \cite{benztrade}. Moreover, \cite{nanda2021fairness} and \cite{tian2021analysis} argue that traditional notions of fairness are not sufficient when the model is vulnerable to adversarial attacks, investigate the class-wise robustness and propose methods to improve the robustness of the most vulnerable class, so as to obtain a fairer robust model.

\section{Attack with one adversarial data point} \label{sec:one point}
In this section, we consider the scenario where the attacker can add one carefully designed adversarial data point to the existing dataset.
\subsection{Problem formulation}
Using a set of training samples $\{\bm{x}_i,y_i,G_i\}_{i=1}^n \coloneqq \{ \bm{X},\bm{y},\bm{G}\}$, where $\bm{x}_i \in \mathbb{R}^{p}$ is the feature vector, $y_i$ is the response variable and $G_i$ indicates the group membership or sensitive status (for example, race, gender), we aim to develop a model that can predict the value of a target variable $Y$ from the input variables $X$. In this paper, we consider the case when there are only two groups, i.e., $G_i \in \{1,2\}$ and assume that the first $m$ training samples are from group $1$ and the remaining samples are from group $2$. For simplification, we denote $\bm{X}=[\bm{X}_1;\bm{X}_2], \bm{y}=[\bm{y}_1;\bm{y}_2]$. 

To build a robust model, we assume that there is an adversary who can observe the whole training dataset and then carefully design an adversarial data point, $\{\bm{x}_0,y_0,G_0\}$, and add it into the existing dataset. After inserting this poisoned data point, we have the poisoned dataset $\{\hat{\bm{X}},\hat{\bm{y}},\hat{\bm{G}}\}$, where $\hat{\bm{ X}} = [\bm{x}_0, \bm{x}_1,\cdots,\bm{x}_n]^T, \hat{\bm{y}} = [y_0, y_1,\cdots,y_n]^T,\hat{\bm{G}} = [G_0, G_1,\cdots,G_n]^T$.
From this poisoned dataset, we aim to design a robust fairness-aware regression model.

In order to characterize both prediction and fairness performance, we consider the following objective function
\begin{eqnarray}
	L=f(\bm{\beta},\hat{\bm{X}},\hat{\bm{y}},\hat{\bm{G}})+ \lambda F(\bm{\beta},\hat{\bm{X}},\hat{\bm{y}},\hat{\bm{G}}), \label{eq:obj func}
\end{eqnarray}
where $\bm{\beta}$ is the regression coefficient, $f(\bm{\beta},\hat{\bm{X}},\hat{\bm{y}},\hat{\bm{G}})$ corresponds to the prediction accuracy loss, $F(\bm{\beta},\hat{\bm{X}},\hat{\bm{y}},\hat{\bm{G}})$ corresponds to the group fairness gap and $\lambda$ is the trade-off parameter. The goal of the adversary is to maximize \eqref{eq:obj func} to make the model less fair and less accurate, while the robust fairness-aware regression model aims at minimizing \eqref{eq:obj func}. To make the problem meaningful, we introduce an energy constraint on the adversarial data point and use $\ell_2$ norm to measure the energy. Thus, we have the minimax problem
\begin{eqnarray}
	\min_{\bm{\beta}} \max_{\substack{(\bm{x}_0,y_0,G_0),\\ \text{s.t. } \|[\bm{x}_0^T,y_0]\|_2\leq\eta }}\hspace{-3mm}&& \hspace{-5mm}L=f(\bm{\beta},\hat{\bm{X}},\hat{\bm{y}},\hat{\bm{G}})+ \lambda F(\bm{\beta},\hat{\bm{X}},\hat{\bm{y}},\hat{\bm{G}}). \nonumber\\
	\label{eq:minimax} 
\end{eqnarray}

To measure the prediction accuracy, we consider the mean-squared error (MSE),
\begin{eqnarray}
	f(\bm{\beta},\hat{\bm{X}},\hat{\bm{y}},\hat{\bm{G}})=\mathbb{E}[(Y-\hat{Y} )^2],\nonumber
\end{eqnarray}
where $\hat{Y}$ is the prediction result. For the group fairness gap, we consider a measure that is closely related to the accuracy parity criterion \cite{chi2021understanding}, $$\mathbb{E}[(Y-\hat{Y})^2 \vert G=1]=\mathbb{E}[(Y-\hat{Y})^2 \vert G=2].$$ Then the absolute difference between two groups can be used to measure the severity of violations \cite{shah2021selective} and we have $$F(\bm{\beta},\hat{\bm{X}},\hat{\bm{y}},\hat{\bm{G}})=\vert\mathbb{E}[(Y-\hat{Y})^2 \vert G=1]-\mathbb{E}[(Y-\hat{Y})^2 \vert G=2]\vert. $$

\subsection{Proposed method}
To solve the minimax problem in \eqref{eq:minimax}, we will first solve the inner maximization problem with respect to the adversary to design the optimal adversarial data point $\{\bm{x}_0,y_0,G_0\}$ under the energy constraint. Then we will solve the outer minimization problem to find a robust fairness-aware model that can optimize both prediction accuracy and the group fairness guarantee.

\noindent\textbf{Maximization Problem} \\
In the following, we want to find the optimal $\{\bm{x}_0,y_0,G_0\}$ for any given $\bm{\beta}$. We first note that there are two choices of $G_0$, and the form of the objective function $L$ under different choices of $G_0$ is different. For $G_0=1$, the objective function $L$ can be written as 
\begin{eqnarray}
	&&\hspace{-12mm}L_1=\frac{1}{n+1}\left(\|y_0-\bm{x}_0^T \bm{\beta}\|_2^2+\|\bm{y_1}-\bm{X}_1\bm{\beta}\|_2^2\right. \nonumber\\
	&&\hspace{-10mm}\left. +\|\bm{y_2}-\bm{X}_2\bm{\beta}\|_2^2\right) +\lambda \left|\frac{1}{m+1}\|y_0-\bm{x}_0^T \bm{\beta}\|_2^2 \right.\nonumber\\
	&&\hspace{-10mm}\left.+\frac{1}{m+1}\|\bm{y_1}-\bm{X}_1\bm{\beta}\|_2^2  -\frac{1}{n-m}\|\bm{y_2}-\bm{X}_2\bm{\beta}\|_2^2\right|.\nonumber
\end{eqnarray}
For $G_0=2$, the objective function $L$ can be written as 
\begin{eqnarray}
	&&\hspace{-12mm}L_2=\frac{1}{n+1}\left(\|y_0-\bm{x}_0^T \bm{\beta}\|_2^2+\|\bm{y_1}-\bm{X}_1\bm{\beta}\|_2^2\right. \nonumber\\
	&&\hspace{-10mm}\left.  +\|\bm{y_2}-\bm{X}_2\bm{\beta}\|_2^2\right)+\lambda \left|\frac{1}{m}\|\bm{y_1}-\bm{X}_1 \bm{\beta}\|_2^2 \right. \nonumber\\
	&&\hspace{-10mm}\left. -\frac{1}{n-m+1}\|{y_0}-\bm{x}_0^T\bm{\beta}\|_2^2 -\frac{1}{n-m+1}\|\bm{y_2}-\bm{X}_2\bm{\beta}\|_2^2\right|.\nonumber
\end{eqnarray}
It is worth noting that for either $L_1$ or $L_2$, the objective function of the minimax problem \eqref{eq:minimax} is non-smooth noncovex-nonconcave. However, we observe that by exploring four different cases 
depending on the value of $G_0$ and the signs of the terms inside $|\cdot|$, the maximization problem can be solved exactly as shown in the following theorem.
\begin{thm} \label{thm:L upper-bound}
	For any given $\bm{\beta}$, we have
	\begin{eqnarray} \max_{\substack{(\bm{x}_0,y_0,G_0),\\\text{s.t. } \|[\bm{x}_0^T,y_0]\|_2\leq\eta }}\hspace{-3mm}&&  L
		\overset{(a)}{=} \max\{g_1(\bm{\beta}),h_1(\bm{\beta}),g_2(\bm{\beta}),h_2(\bm{\beta})\}, \nonumber
	\end{eqnarray}
	where 
	\begin{eqnarray}
		g_1(\bm{\beta})&=&C_{g_1}\eta^2(1+\|\bm{\beta}\|_2^2)  +C_{g_1}\|\bm{y_1}-\bm{X}_1 \bm{\beta}\|_2^2  \nonumber\\
		&&+D_{g_1}\|\bm{y_2}-\bm{X}_2 \bm{\beta}\|_2^2, \nonumber \\
		h_1(\bm{\beta})&=&\max\{0,C_{h_1}\} \eta^2(1+\|\bm{\beta}\|_2^2) +C_{h_1}\|\bm{y_1}-\bm{X}_1 \bm{\beta}\|_2^2  \nonumber\\
		&&+D_{h_1}\|\bm{y_2}-\bm{X}_2 \bm{\beta}\|_2^2, \nonumber \\
		g_2(\bm{\beta})&=&\max\{0,D_{g_2}\}
		\eta^2(1+\|\bm{\beta}\|_2^2)  +C_{g_2}\|\bm{y_1}-\bm{X}_1 \bm{\beta}\|_2^2  \nonumber\\
		&&+D_{g_2}\|\bm{y_2}-\bm{X}_2 \bm{\beta}\|_2^2, \nonumber\\
		h_2(\bm{\beta})&=&D_{h_2} \eta^2(1+\|\bm{\beta}\|_2^2) +C_{h_2}\|\bm{y_1}-\bm{X}_1 \bm{\beta}\|_2^2  \nonumber\\
		&&+D_{h_2}\|\bm{y_2}-\bm{X}_2 \bm{\beta}\|_2^2, \nonumber
	\end{eqnarray}
	with $C_{g_1}=\frac{\lambda}{m+1}+\frac{1}{n+1},D_{g_1}=-\frac{\lambda}{n-m}+\frac{1}{n+1},
	C_{h_1}=-\frac{\lambda}{m+1}+\frac{1}{n+1},D_{h_1}=\frac{\lambda}{n-m}+\frac{1}{n+1},
	C_{g_2}=\frac{\lambda}{m}+\frac{1}{n+1},D_{g_2}=-\frac{\lambda}{n-m+1}+\frac{1}{n+1},
	C_{h_2}=-\frac{\lambda}{m}+\frac{1}{n+1},D_{h_2}=\frac{\lambda}{n-m+1}+\frac{1}{n+1}. $
	Denote $\tilde{\bm{x}}_0=[{\bm{x}}_0^T,{y}_0]^T, \bm{b}=[\bm{\beta}^T,-1]^T$. Then we have
	\begin{itemize}
		\item when either of the following occurs: 1) $g_1(\bm{\beta})\geq \max\{h_1(\bm{\beta}),g_2(\bm{\beta}),h_2(\bm{\beta})\}$, 2) $h_1(\bm{\beta})\geq \max\{g_1(\bm{\beta}),$ $g_2(\bm{\beta}),h_2(\bm{\beta})\} \text{ and } C_{h_1}\geq 0$, the maximum value of $L$ (equality (a)) is achieved if $\tilde{\bm{x}}_0^*(\bm{\beta})=\eta \frac{\bm{b}}{\|\bm{b}\|_2}$ and $G_0=1$;
		\item when $ h_1(\bm{\beta}) \geq \max\{g_1(\bm{\beta}),g_2(\bm{\beta}),h_2(\bm{\beta})\}\text{ and } C_{h_1} < 0$, (a) is attained as long as $\tilde{\bm{x}}_0^*(\bm{\beta}) \perp \bm{b}$ and $G_0=1$;
		\item when either of the following occurs: 1) $g_2(\bm{\beta}) \geq \max\{g_1(\bm{\beta}),h_1(\bm{\beta}),h_2(\bm{\beta})\}\text{ and } D_{g_2}\geq 0$, 2) $h_2(\bm{\beta})\geq \max\{g_1(\bm{\beta}),h_1(\bm{\beta}),g_2(\bm{\beta})\}$, (a) is attained if $\tilde{\bm{x}}_0^*(\bm{\beta})=\eta \frac{\bm{b}}{\|\bm{b}\|_2}$ and $G_0=2$;
		\item when $g_2(\bm{\beta}) \geq \max\{g_1(\bm{\beta}),h_1(\bm{\beta}),h_2(\bm{\beta})\}\text{ and } D_{g_2} < 0$, (a) is attained if $\tilde{\bm{x}}_0^*(\bm{\beta}) \perp \bm{b}$ and $G_0=2$. 
	\end{itemize}
	
\end{thm}

\begin{IEEEproof}
	Please refer to Appendix~\ref{app:L upper-bound}.
\end{IEEEproof}

\begin{rmk} \label{rmk:involve beta}
	$g_1(\bm{\beta})$, $h_1(\bm{\beta})$, $g_2(\bm{\beta})$ and $h_2(\bm{\beta})$ involve $\bm{\beta}$ only through $\|\bm{\beta}\|_2^2$, $ \|\bm{y_1}-\bm{X}_1 \bm{\beta}\|_2^2$ and  $\|\bm{y_2}-\bm{X}_2 \bm{\beta}\|_2^2$. 
	Furthermore, from Theorem~\ref{thm:L upper-bound}, for $G_0=1$, we have
	$$
	\max \limits_{\substack{(\bm{x}_0,y_0,1),s.t. \|[\bm{x}_0^T,y_0]\|_2\leq\eta }} L_1{=} \max\{g_1(\bm{\beta}),h_1(\bm{\beta})\},$$
	where $g_1(\bm{\beta})$ corresponds to the case in which the terms inside $|\cdot|$ of $L_1$ is non-negative and $h_1(\bm{\beta})$ corresponds to the case in which the terms inside $|\cdot|$ is negative. Subsequently, for the conditions of equality, we discuss two cases $L_1=g_1(\bm{\beta}) \geq h_1(\bm{\beta})$ and $L_1=h_1(\bm{\beta})>g_1(\bm{\beta})$, where there are two sub-cases for $L_1=h_1(\bm{\beta})$ based on the value of $C_{h_1}$. There are similar observations for $G_0=2$.
\end{rmk}

\noindent\textbf{Minimization Problem} \\
Using Theorem~\ref{thm:L upper-bound}, the original minmax problem is converted to the following problem
\begin{eqnarray}
	\min_{\bm{\beta}} \max_{(\bm{x}_0,y_0,G_0)} L = \min_{\bm{\beta}} \max\{g_1(\bm{\beta}),h_1(\bm{\beta}),g_2(\bm{\beta}),h_2(\bm{\beta})\}. \label{eq:minmax gh}
\end{eqnarray}
As we seek to minimize the largest of four functions, \eqref{eq:minmax gh} can be separated into four sub-problems. One of them is 
\begin{eqnarray}
	\min_{\bm{\beta}}&& \hspace{-5mm} g_1(\bm{\beta}), \nonumber\\
	\text{s.t. } && \hspace{-5mm}g_1(\bm{\beta}) \geq g_2(\bm{\beta}), g_1(\bm{\beta}) \geq h_1(\bm{\beta}), g_1(\bm{\beta}) \geq h_2(\bm{\beta}), \label{eq:minimization g1 simple}
\end{eqnarray}
and other sub-problems can be written in a similar manner. Once these sub-problems are solved, the solution to \eqref{eq:minmax gh} can be obtained.
	
	For notation simplicity, we denote 
$\frac{1}{2}\frac{\partial^2 g_1(\bm{\beta})}{\partial \bm{\beta}^2} = C_{g_1}(\eta^2\bm{I}+  \bm{X}_1^T\bm{X}_1)+D_{g_1}\bm{X}_2^T\bm{X}_2 
		\coloneqq \bm{M_{g_1}}$, 
$\frac{1}{2}\frac{\partial^2 h_1(\bm{\beta})}{\partial \bm{\beta}^2} =\max\left\{0,C_{h_1}\right\}\eta^2\bm{I}+ C_{h_1} \bm{X}_1^T\bm{X}_1
		+D_{h_1}\bm{X}_2^T\bm{X}_2\coloneqq \bm{M_{h_1}}$, 
$\frac{1}{2}\frac{\partial^2 g_2(\bm{\beta})}{\partial \bm{\beta}^2} = \max\left\{0,D_{g_2}\right\}\eta^2\bm{I}+ C_{g_2} \bm{X}_1^T\bm{X}_1
		+D_{g_2}\bm{X}_2^T\bm{X}_2
		\coloneqq \bm{M_{g_2}}$,
$\frac{1}{2}\frac{\partial^2 h_2(\bm{\beta})}{\partial \bm{\beta}^2} = C_{h_2}\bm{X}_1^T\bm{X}_1+D_{h_2}(\eta^2\bm{I}+  \bm{X}_2^T\bm{X}_2)
		\coloneqq \bm{M_{h_2}}$.

	In the following, we focus on solving \eqref{eq:minimization g1 simple}. The analysis of other sub-problems can be done similarly. Specifically, \eqref{eq:minimization g1 simple} can be further written as
	\begin{eqnarray}
		\min_{\bm{\beta}}&&\hspace{-3mm} g_1(\bm{\beta})=C_{g_1}\eta^2(1+\|\bm{\beta}\|_2^2)  +C_{g_1}\|\bm{y_1}-\bm{X}_1 \bm{\beta}\|_2^2 \nonumber\\
		&&\hspace{10mm}+D_{g_1}\|\bm{y_2}-\bm{X}_2 \bm{\beta}\|_2^2,  \label{eq:minimization g1}\\
		\text{s.t. }&& 
		C_{1}(\bm{\beta}) =g_1(\bm{\beta})- h_1(\bm{\beta}) \geq 0,  \nonumber\\
		&&C_{2}(\bm{\beta}) =g_1(\bm{\beta})- g_2(\bm{\beta}) \geq 0, \nonumber\\
		&& C_{3}(\bm{\beta}) =g_1(\bm{\beta})- h_2(\bm{\beta}) \geq 0.  
		\label{eq:constraint}
	\end{eqnarray}
	For the objective function in \eqref{eq:minimization g1}, since $D_{g_1}$ can be negative, $\bm{M_{g_1}}$ is not necessarily positive-semidefinite. Hence, \eqref{eq:minimization g1} is a non-convex quadratic minimization problem with several quadratic constraints (QCQP), which is NP hard in general \cite{luo2007approximation}. Despite this challenge, we are able to solve this problem by exploiting the structure inherent to our problem. The following proposition gives us sufficient conditions for global minimizers of QCQP, following from \textit{Proposition 3.2} in \cite{jeyakumar2007non}. 
	\begin{prop} 
		\label{prop:global minimizer}
		If $\exists \alpha_i \geq 0, i=1,2,3$ such that for $\bm{\beta}=\bm{\beta}^*$,
		\begin{eqnarray}
			&& \bm{M_{g_1}}-\sum_{i=1}^3 \alpha_i \frac{\partial^2 C_i(\bm{\beta})}{\partial \bm{\beta}^2}\succeq \bm{0}, \nonumber\\
			&& \frac{\partial g_1(\bm{\beta})}{\partial \bm{\beta}} \vert_{\bm{\beta}^*}-\sum_{i=1}^3 \alpha_i \frac{\partial C_i(\bm{\beta})}{\partial \bm{\beta}}\vert_{\bm{\beta}^*} = \bm{0},\nonumber\\
			&&\sum_{i=1}^3 \alpha_i C_i(\bm{\beta}^*) = 0, \label{eq:kkt} \\
			&& C_i(\bm{\beta}^*) \geq 0, i=1,2,3, \nonumber
		\end{eqnarray}
		then $\bm{\beta}^*$ is a global minimizer of QCQP \eqref{eq:minimization g1}. 
	\end{prop}

	\begin{rmk}
		From \eqref{eq:kkt}, we have that for each constraint $C_i(\bm{\beta})$, there are two possible cases: 1) $\alpha_i=0, C_i(\bm{\beta}^*) \geq 0$; 2) $\alpha_i>0, C_i(\bm{\beta}^*) = 0$. In total, there will be $2^3$ cases of different combinations of $\alpha_i$s. By examining these $8$ different cases, we can obtain the optimal regression coefficient $\bm{\beta}^*$ of the sub-problem \eqref{eq:minimization g1}.
	\end{rmk}
	
	In the following, we will analyze four types of cases sequentially: 1) $\alpha_1=\alpha_2=\alpha_3 = 0$; 2) the case with only one non-zero $\alpha_i$, i.e. $ \exists ! \alpha_i > 0$ and $\alpha_k=0, \forall k \neq i$; 3) the case with two non-zero $\alpha_i$s, i.e. $\exists i, j, i \neq j, \alpha_i > 0, \alpha_j > 0$ and $\alpha_k=0, \forall k \notin \{i,j\}$; 4) $\alpha_i> 0, i=1,2,3$. 

	\textit{Case 1: $\alpha_1=\alpha_2=\alpha_3 = 0$}\\
	By Proposition~\ref{prop:global minimizer}, if there exists $\tilde{\bm{\beta}}$, such that
	\begin{eqnarray}
		&&\bm{M_{g_1}}\succeq \bm{0},\label{eq:sec-order 1}\\
		&&\frac{\partial g_1(\bm{\beta})}{\partial \bm{\beta}} \vert_{\tilde{\bm{\beta}}} = \bm{0},\label{eq:fir-order 1}
	\end{eqnarray}
	\begin{eqnarray}
		&& C_i(\tilde{\bm{\beta}}) \geq 0, i=1,2,3, \label{eq:opt constraint 1}
	\end{eqnarray}
	then $\tilde{\bm{\beta}}$ is a global minimizer of \eqref{eq:minimization g1}. From \eqref{eq:sec-order 1}, we require that $\bm{M_{g_1}}$ is positive-semidefinite, which can be true when $\lambda$ is small, e.g. when $D_{g_1}\geq 0$. From \eqref{eq:fir-order 1}, when $\bm{M_{g_1}}$ is invertible, we have 
	\begin{eqnarray}
		\tilde{\bm{\beta}} = \bm{M_{g_1}}^{-1}\left[ C_{g_1} \bm{X}_1^T\bm{Y}_1+D_{g_1}\bm{X}_2^T\bm{Y}_2 \right]. \label{eq:beta solution 1}
	\end{eqnarray}
	If \eqref{eq:opt constraint 1} is satisfied at \eqref{eq:beta solution 1}, then $\tilde{\bm{\beta}} $ is a global minimizer of \eqref{eq:minimization g1}. Otherwise, there does not exist a global minimizer in \textit{Case 1} and we will consider \textit{Case 2}.

	\textit{Case 2: $\exists ! \alpha_i > 0$ and  $\alpha_k=0, \forall k \neq i$}\\
	We will consider the particular case $\alpha_1 > 0,  \alpha_2=\alpha_3=0$ and other cases can be analyzed similarly. 
	
	By Proposition~\ref{prop:global minimizer}, if there exists $\bar{\bm{\beta}}$ and $\alpha_1 > 0$, such that
	\begin{eqnarray}
		&& \bm{M_{g_1}}-\alpha_1 ( \bm{M_{g_1}}-\bm{M_{h_1}})\succeq 0,\label{eq:sec-order 2}\\
		&&\frac{\partial g_1(\bm{\beta})}{\partial \bm{\beta}} \vert_{\bar{\bm{\beta}}}-\alpha_1 \frac{\partial C_1(\bm{\beta})}{\partial \bm{\beta}}\vert_{\bar{\bm{\beta}}} = \bm{0},\label{eq:fir-order 2}\\
		&& C_1(\bar{\bm{\beta}}) = 0,  
		 C_2(\bar{\bm{\beta}}) \geq 0, C_3(\bar{\bm{\beta}}) \geq 0,\label{eq:opt ineq constraint 2}
	\end{eqnarray}
	then $\bar{\bm{\beta}}$ is a global minimizer of \eqref{eq:minimization g1}.

	\begin{prop} \label{prop:case 2}
		Denote the largest eigenvalue of $\bm{X}_1^T\bm{X}_1$ as $v_{X_1,p}$ and the largest eigenvalue of $\bm{X}_2^T\bm{X}_2$ as $v_{X_2,p}$.  Assuming that $\eta^2 \geq \eta_{\min}^2= \max\left\{\frac{(n+1)v_{X_1,p}}{m(m+1)},\frac{(n+1)v_{X_2,p}}{(n-m+1)(n-m)}\right\}$, we have $A_{g_1h_1}=\{ \alpha: \bm{M_{g_1}}-\alpha ( \bm{M_{g_1}}-\bm{M_{h_1}})\succ 0\} \neq \emptyset$. By randomly selecting an $\alpha_1^* \in A_{g_1h_1}$, for 	
		\begin{eqnarray}
			\check{{\bm{\beta}}}&=&\left[(1-\alpha_1^*-\gamma^*)\bm{M}_{g_1}+(\alpha_1^*+\gamma^*) \bm{M}_{h_1}\right]^{-1} \nonumber\\
			&&\cdot \left[(1-\alpha_1^*-\gamma^*)\bm{E}_{g_1}-(\alpha_1^*+\gamma^*) \bm{E}_{h_1}\right],\nonumber
		\end{eqnarray} 
	where $\gamma^*$ is a certain Lagrangian multiplier, and $\bm{E}_{g_1}={C}_{g_1}\bm{X}_1^T\bm{y_1} +{D}_{g_1}\bm{X}_2^T\bm{y_2}, ~
	\bm{E}_{h_1}={C}_{h_1}\bm{X}_1^T\bm{y_1}+{D}_{h_1}\bm{X}_2^T\bm{y_2}$, if we have $C_2(\check{\bm{\beta}}) \geq 0, C_3(\check{\bm{\beta}}) \geq 0$, then $ \check{\bm{\beta}}$ satisfies \eqref{eq:sec-order 2}, \eqref{eq:fir-order 2}, \eqref{eq:opt ineq constraint 2} and is a global minimizer of \eqref{eq:minimization g1}. 
	\end{prop}
\begin{IEEEproof}
	Please refer to Appendix~\ref{app:case 2 analysis}.
	\end{IEEEproof}

	\textit{Case 3: $\exists i, j, i \neq j, \alpha_i > 0, \alpha_j > 0$ and $\alpha_k=0, \forall k \notin \{i,j\}$}\\
	We will consider the particular case $\alpha_1 > 0,  \alpha_2>0, \alpha_3=0$ and other cases can be analyzed in a similar manner. By Proposition~\ref{prop:global minimizer}, if there exists $\hat{\bm{\beta}}$ and $\alpha_1 > 0, \alpha_2>0$, such that
	\begin{eqnarray}
		&& \hspace{-5mm}\bm{M_{g_1}}-\alpha_1 ( \bm{M_{g_1}}-\bm{M_{h_1}})-\alpha_2 ( \bm{M_{g_1}}-\bm{M_{g_2}})\succeq 0,\label{eq:sec-order double}\\
		&& \hspace{-5mm}\frac{\partial g_1(\bm{\beta})}{\partial \bm{\beta}} \vert_{\hat{\bm{\beta}}}-\alpha_1 \frac{\partial C_1(\bm{\beta})}{\partial \bm{\beta}}\vert_{\hat{\bm{\beta}}}-\alpha_2 \frac{\partial C_2(\bm{\beta})}{\partial \bm{\beta}}\vert_{\hat{\bm{\beta}}} = \bm{0},\label{eq:fir-order double}\\
		&&  \hspace{-5mm}C_1(\hat{\bm{\beta}}) = 0, C_2(\hat{\bm{\beta}}) = 0, \label{eq:opt constraint double}\\
		&&   \hspace{-5mm} C_3(\hat{\bm{\beta}}) \geq 0, \label{eq:opt ineq constraint double}
	\end{eqnarray}
	then $\hat{\bm{\beta}}$ is a global minimizer of \eqref{eq:minimization g1}.

	\begin{prop} \label{prop:case 3}
		For
		\begin{eqnarray}
			&& \hspace{-7mm}\check{{\bm{\beta}}}=\left[(1-\alpha_1^*-\gamma_1^*-\gamma_2^*)\bm{M}_{g_1} +(\alpha_1^*+\gamma_1^*)\bm{M}_{h_1}+\gamma_2^*\bm{M}_{g_2}\right]^{-1}\nonumber\\
			&&\hspace{-2mm}\cdot\left[(1-\alpha_1^*-\gamma_1^*-\gamma_2^*)\bm{E}_{g_1}+(\alpha_1^*+\gamma_1^*)\bm{E}_{h_1}+\gamma_2^*\bm{E}_{g_2}\right], \nonumber
		\end{eqnarray}
	where $\gamma_1^*,\gamma_2^*$ are certain Lagrangian multipliers, and $\bm{E}_{g_2}=C_{g_2}\bm{X}_1^T\bm{y_1} +D_{g_2}\bm{X}_2^T\bm{y_2}$,
	if $C_3(\check{\bm{\beta}}) \geq 0$, then $ \check{\bm{\beta}}$ satisfies \eqref{eq:sec-order double}, \eqref{eq:fir-order double}, \eqref{eq:opt constraint double}, \eqref{eq:opt ineq constraint double} and is a global minimizer of \eqref{eq:minimization g1}. 
	\end{prop}
\begin{IEEEproof}
	Please refer to Appendix~\ref{app:case 3 analysis}. 
	\end{IEEEproof}

	\textit{Case 4: $\alpha_i>0, i=1,2,3$}\\
	By Proposition~\ref{prop:global minimizer}, if there exists $\acute{\bm{\beta}}$ and $\alpha_i > 0,i=1,2,3$, such that
	\begin{eqnarray}
		&& \bm{M_{g_1}}-\alpha_1 ( \bm{M_{g_1}}-\bm{M_{h_1}})-\alpha_2 ( \bm{M_{g_1}}-\bm{M_{g_2}})\nonumber\\
		&&\hspace{33mm}-\alpha_3 ( \bm{M_{g_1}}-\bm{M_{h_2}})\succeq 0,\label{eq:sec-order tri}\\
		&&\frac{\partial g_1(\bm{\beta})}{\partial \bm{\beta}} \vert_{\acute{\bm{\beta}}}-\alpha_1 \frac{\partial C_1(\bm{\beta})}{\partial \bm{\beta}}\vert_{\acute{\bm{\beta}}}-\alpha_2 \frac{\partial C_2(\bm{\beta})}{\partial \bm{\beta}}\vert_{\acute{\bm{\beta}}}\nonumber\\
		&&\hspace{33mm}-\alpha_3 \frac{\partial C_3(\bm{\beta})}{\partial \bm{\beta}}\vert_{\acute{\bm{\beta}}} = \bm{0},\label{eq:fir-order tri}\\
		&& C_1(\acute{\bm{\beta}}) = 0, C_2(\acute{\bm{\beta}}) = 0, C_3(\acute{\bm{\beta}}) = 0, \label{eq:opt constraint tri}
	\end{eqnarray}
	then $\acute{\bm{\beta}}$ is a global minimizer of \eqref{eq:minimization g1}. From Remark~\ref{rmk:involve beta}, we note that with \eqref{eq:opt constraint tri}, there are three equations on $\|\bm{\beta}\|_2^2$, $ \|\bm{y_1}-\bm{X}_1 \bm{\beta}\|_2^2$ and  $\|\bm{y_2}-\bm{X}_2 \bm{\beta}\|_2^2$, which indicates that there will be deterministic solutions for them or the feasible set is empty. 
	
	When the feasible set of \eqref{eq:opt constraint tri} is nonempty (for example, when $\lambda > \max\{\frac{m+1}{n+1},\frac{n-m+1}{n+1} \}$), the value of $g_1(\bm{\beta}), C_1(\bm{\beta}), C_2(\bm{\beta}), C_3(\bm{\beta})$ is determined as there have been deterministic solutions for $\|\bm{\beta}\|_2^2$, $ \|\bm{y_1}-\bm{X}_1 \bm{\beta}\|_2^2$ and  $\|\bm{y_2}-\bm{X}_2 \bm{\beta}\|_2^2$. Then the process of finding $\acute{\bm{\beta}}$ is \\
	\noindent 1. Solve \eqref{eq:opt constraint tri} and derive the solution for $\|\bm{\beta}\|_2^2$, $ \|\bm{y_1}-\bm{X}_1 \bm{\beta}\|_2^2$ and  $\|\bm{y_2}-\bm{X}_2 \bm{\beta}\|_2^2$.\\
	\noindent 2. Calculate the value of $g_1(\bm{\beta}), C_1(\bm{\beta}), C_2(\bm{\beta}), C_3(\bm{\beta})$. \\
	\noindent 3. Select $\alpha_1,\alpha_2,\alpha_3$ such that \eqref{eq:sec-order tri} is satisfied. Then \eqref{eq:fir-order tri} is satisfied naturally as $g_1(\bm{\beta}), C_1(\bm{\beta}), C_2(\bm{\beta}), C_3(\bm{\beta})$ are constants.

\section{Rank-one attack} \label{sec:rank-one}
In Section \ref{sec:one point}, we have discussed how to design one adversarial point to attack the fair regression model. In this section, we consider a more powerful adversary who can observe the whole training dataset and then perform a rank-one attack on the feature matrix. This type of attack covers many practical scenarios, for example, modifying one entry of the feature matrix, deleting one feature, changing one feature, replacing one feature, etc~\cite{9024000}. In particular, the attacker will carefully design a rank-one feature modification matrix $\bm{\Delta}$ and add it to the original feature matrix $\bm{X}$, so as to obtain the modified feature matrix $\hat{\bm{X}}=\bm{X}+\bm{\Delta}$. Since $\bm{\Delta}$ is of rank 1, we can write $\bm{\Delta}=\bm{c}\bm{d}^T$, where $\bm{c} \in \mathbb{R}^n$ and $\bm{d} \in \mathbb{R}^p$. Moreover, recall that there are samples from two groups, we denote the modification matrix of the first group as $\bm{\Delta}_1$, i.e., the first $m$ rows of $\bm{\Delta}$, and assume that $\bm{\Delta}_1=\bm{c}_1 \bm{d}^T$, where $\bm{c}_1$ consists of the first $m$ components of $\bm{c}$. Similarly, for the second group, the modification matrix is $\bm{\Delta}_2=\bm{c}_2 \bm{d}^T$. Then the modified feature matrices for two groups are $\hat{\bm{X}}_1=\bm{X}_1+\bm{\Delta}_1$ and $\hat{\bm{X}}_2=\bm{X}_2+\bm{\Delta}_2$.

Similar to Section~\ref{sec:one point}, we introduce an energy constraint on the rank-one attack. We use the Frobenius norm to measure the energy of the modification matrix $\bm{\Delta}$. Recall that $\bm{y}, \bm{G}$ remain unchanged in this attack scheme, we have the minimax problem
\begin{eqnarray}
	\min_{\bm{\beta}} \max_{\|{\Delta}\|_F\leq\eta }&& \hspace{-5mm}f(\bm{\beta},\hat{\bm{X}})+ \lambda F(\bm{\beta},\hat{\bm{X}}). \label{eq:minimax rank} 
\end{eqnarray}

To solve \eqref{eq:minimax rank}, we will first investigate the inner maximization problem. We will perform various variable augmentations, and convert the maximization problem into a form with five arguments, four of which can be solved exactly. Then we will transform the original nonconvex-nonconcave minimax problem into several weakly-convex-weakly-concave minimax problems.
 
\noindent\textbf{Maximization problem}\\
For the objective function in \eqref{eq:minimax rank}, we have
\begin{eqnarray}
	&& \hspace{-7mm}f(\bm{\beta},\hat{\bm{X}})+ \lambda F(\bm{\beta},\hat{\bm{X}}) 
	=\frac{1}{n} \|\bm{y}-\bm{\hat{X}}\bm{\beta}\|_2^2\nonumber\\
	&& +\lambda \left| \frac{1}{m} \|\bm{y}_1-\bm{\hat{X}}_1\bm{\beta}\|_2^2-\frac{1}{n-m} \|\bm{y}_2-\bm{\hat{X}}_2\bm{\beta}\|_2^2\right| \nonumber\\
	&&\hspace{-11mm}=\max\{g(\bm{\beta},\bm{\hat{X}}),h(\bm{\beta},\bm{\hat{X}})\}, \nonumber
\end{eqnarray}
in which 
$$g(\bm{\beta},\bm{\hat{X}})= C_g \|\bm{y}_1-\bm{\hat{X}}_1\bm{\beta}\|_2^2+D_g \|\bm{y}_2-\bm{\hat{X}}_2\bm{\beta}\|_2^2,$$ 
$$h(\bm{\beta},\bm{\hat{X}})=C_h \|\bm{y}_1-\bm{\hat{X}}_1\bm{\beta}\|_2^2+D_h \|\bm{y}_2-\bm{\hat{X}}_2\bm{\beta}\|_2^2,$$
with $C_g=\frac{1}{n}+\frac{\lambda}{m}$, $D_g=\frac{1}{n}-\frac{\lambda}{n-m}$, $C_h=\frac{1}{n}-\frac{\lambda}{m}$, $D_h=\frac{1}{n}+\frac{\lambda}{n-m}$.

\begin{lem} \label{lm:convexity of c and d two group}
	For $g(\bm{\beta},\bm{\hat{X}})$ and $h(\bm{\beta},\bm{\hat{X}})$, we have that
	\begin{enumerate}
		\item if $D_g \geq 0$, $g(\bm{\beta},\bm{\hat{X}})$ is convex in $\bm{c}_1$ for any given $\bm{c}_2,\bm{d}$, and also convex in $\bm{c}_2$ for any given $\bm{c}_1,\bm{d}$; otherwise, $g(\bm{\beta},\bm{\hat{X}})$ is convex in $\bm{c}_1$ for any given $\bm{c}_2,\bm{d}$, and concave in $\bm{c}_2$ for any given $\bm{c}_1,\bm{d}$;
		\item if $C_h \geq 0$, $h(\bm{\beta},\bm{\hat{X}})$ is convex in $\bm{c}_1$ for any given $\bm{c}_2,\bm{d}$, and also convex in $\bm{c}_2$ for any given $\bm{c}_1,\bm{d}$; otherwise, $h(\bm{\beta},\bm{\hat{X}})$ is concave in $\bm{c}_1$ for any given $\bm{c}_2,\bm{d}$, and convex in $\bm{c}_2$ for any given $\bm{c}_1,\bm{d}$. 
	\end{enumerate}
\end{lem}

Based on Lemma~\ref{lm:convexity of c and d two group}, we now solve the maximization problem in \eqref{eq:minimax rank}. First, note that 
\begin{eqnarray}
	&&\hspace{-8mm}\max_{\|\bm{c}\bm{d}^T \|_F \leq \eta} \max\{g(\bm{\beta},\bm{\hat{X}}),h(\bm{\beta},\bm{\hat{X}})\} \nonumber\\
	&=& \max\left\{\max_{\|\bm{c}\bm{d}^T \|_F \leq \eta}g(\bm{\beta},\bm{\hat{X}}),\max_{\|\bm{c}\bm{d}^T \|_F \leq \eta}h(\bm{\beta},\bm{\hat{X}})\right\}, \nonumber
\end{eqnarray}
which indicates that the maximization problem can be separated into two sub-problems. For simplicity of presentation, we will only explore the sub-problem of $g(\bm{\beta},\bm{\hat{X}})$ in detail and the sub-problem of $h(\bm{\beta},\bm{\hat{X}})$ can be analyzed similarly.

\textit{1) Sub-problem of $g(\bm{\beta},\bm{\hat{X}})$}\\
According to Lemma~\ref{lm:convexity of c and d two group}, the value of $D_g$ will affect the property of $g(\bm{\beta},\bm{\hat{X}})$. In the following, we will first explore the case $D_g \geq 0$ and obtain Lemma~\ref{lm:max g D_g positive} as well as Proposition~\ref{prop:max g_a}, and then explore the case $D_g < 0$ and obtain Lemma~\ref{lm:max g D_g negative} as well as Proposition~\ref{prop:max g_b}. 

\begin{lem} \label{lm:max g D_g positive}
	For $D_g \geq 0$, we have
	\begin{eqnarray}
		&&\max_{\|\bm{c}\bm{d}^T \|_F \leq \eta} g(\bm{\beta},\bm{\hat{X}})\nonumber\\
		&&\hspace{-9mm}= \max_{0 < \eta_{c}\leq \eta}\max_{0 \leq \eta_{c_1}\leq \eta_c}\max_{\|\bm{d}\|_2 \leq 1}\max_{\|\bm{c}_2\|_2 = \sqrt{\eta_c^2-\eta_{c_1}^2}} \max_{\|\bm{c}_1\|_2 = \eta_{c_1}} g(\bm{\beta},\bm{\hat{X}}) \nonumber\\
		&&\hspace{-9mm}{=}\max_{0 < \eta_{c}\leq \eta}\max_{0 \leq \eta_{c_1}\leq \eta_c}\max_{\|\bm{d}\|_2 \leq 1} g_{m_1}(\eta_{c_1},\bm{\beta},\bm{d}),\nonumber
	\end{eqnarray}
	where
	\begin{eqnarray}
		g_{m_1}(\eta_{c_1},\bm{\beta},\bm{d})=&C_g(\|\bm{y}_1-\bm{X}_1\bm{\beta}\|_2+\eta_{c_1}\bm{d}^T\bm{\beta})^2\nonumber\\
		&\hspace{-20mm}+D_g(\|\bm{y}_2-\bm{X}_2\bm{\beta}\|_2+\sqrt{\eta_c^2-\eta_{c_1}^2}\bm{d}^T\bm{\beta})^2. \nonumber
	\end{eqnarray}
\end{lem}
\begin{IEEEproof}
	Please refer to Appendix~\ref{app:max g D_g positive}. 
\end{IEEEproof}

Note that $g_{m_1}(\eta_{c_1},\bm{\beta},\bm{d})$ is a quadratic function with respect to $\bm{d}^T\bm{\beta}$, we have the following proposition. 
\begin{prop} \label{prop:max g_a}
	\begin{eqnarray}
	\max_{\|\bm{c}\bm{d}^T \|_F \leq \eta} g(\bm{\beta},\bm{\hat{X}})=
		\max_{0 \leq \eta_{c_1}\leq \eta} g_a(\eta_{c_1},\bm{\beta}),\nonumber
	\end{eqnarray} 
	where $g_a(\eta_{c_1},\bm{\beta})=C_g(\|\bm{y}_1-\bm{X}_1\bm{\beta}\|_2+\eta_{c_1}\|\bm{\beta}\|_2)^2+D_g(\|\bm{y}_2-\bm{X}_2\bm{\beta}\|_2+\sqrt{\eta^2-\eta_{c_1}^2}\|\bm{\beta}\|_2)^2 $. 
\end{prop}
\begin{IEEEproof}
	Please refer to Appendix~\ref{app:max g_a}. 
	\end{IEEEproof}

\begin{lem} \label{lm:max g D_g negative}
	For $D_g < 0$, we have
\begin{eqnarray}
	\max_{\|\bm{c}\bm{d}^T \|_F \leq \eta} g(\bm{\beta},\bm{\hat{X}}){=}\max_{0 < \eta_{c}\leq \eta}\max_{0 \leq \eta_{c_1}\leq \eta_c}\max_{\|\bm{d}\|_2 \leq 1} g_{m_2}(\eta_{c_1},\bm{\beta},\bm{d}),\nonumber
\end{eqnarray}
where
\begin{equation*}
	g_{m_2}(\eta_{c_1},\bm{\beta},\bm{d})=
	\begin{cases}
		C_g(\|\bm{y}_1-\bm{X}_1\bm{\beta}\|_2+\eta_{c_1}\bm{d}^T\bm{\beta})^2, \\
		\hspace{17mm}\text{if } \| \bm{y}_2-\bm{X}_2 \bm{\beta}\|_2 \leq \eta \|\bm{\beta}\|_2,\\
		C_g(\|\bm{y}_1-\bm{X}_1\bm{\beta}\|_2+\eta_{c_1}\bm{d}^T\bm{\beta})^2\\
		\hspace{-1mm}+D_g(\|\bm{y}_2-\bm{X}_2\bm{\beta}\|_2-\sqrt{\eta_c^2-\eta_{c_1}^2}\bm{d}^T\bm{\beta})^2, \\
		\hspace{17mm}\text{ otherwise,}
	\end{cases}
\end{equation*}
\end{lem}
\begin{IEEEproof}
	Please refer to Appendix~\ref{app:max g D_g negative}. 
\end{IEEEproof}

From the above lemma, we have the following proposition. 
\begin{prop} \label{prop:max g_b}
	$$\max \limits_{\|\bm{c}\bm{d}^T \|_F \leq \eta} g(\bm{\beta},\bm{\hat{X}})=\max\limits_{0 \leq \eta_{c_1}\leq \eta}g_b(\eta_{c_1},\bm{\beta}),$$
	where
	\begin{eqnarray}
		&&g_b(\eta_{c_1},\bm{\beta})=
		\begin{cases}
			g_{b_1}(\eta_{c_1},\bm{\beta}), \text{ if }\| \bm{y}_2-\bm{X}_2 \bm{\beta}\|_2 \leq \eta \|\bm{\beta}\|_2,\\
			g_{b_2}(\eta_{c_1},\bm{\beta}), \text{ otherwise. }
		\end{cases} \nonumber\\
	&&g_{b_1}(\eta_{c_1},\bm{\beta})=C_g(\|\bm{y}_1-\bm{X}_1\bm{\beta}\|_2+\eta_{c_1}\|\bm{\beta}\|_2)^2,\nonumber \\
	&&g_{b_2}(\eta_{c_1},\bm{\beta})=\left[C_g(\|\bm{y}_1-\bm{X}_1\bm{\beta}\|_2+\eta_{c_1}\|\bm{\beta}\|_2)^2 \right.\nonumber\\
	&&\hspace{10mm}\left. + D_g(\|\bm{y}_2-\bm{X}_2\bm{\beta}\|_2-\sqrt{\eta^2-\eta_{c_1}}\|\bm{\beta}\|_2)^2\right]. \nonumber
\end{eqnarray}
\end{prop}
\begin{IEEEproof}
	Please refer to Appendix~\ref{app:max g_b}. 
	\end{IEEEproof}

\textit{2) Sub-problem of $h(\bm{\beta},\bm{\hat{X}})$}\\
Following similar process in analyzing the sub-problem of $g(\bm{\beta},\bm{\hat{X}})$, we have that
\begin{itemize}
	\item if $C_h \geq 0$, we have 
	$$\max \limits_{\|\bm{c}\bm{d}^T \|_F \leq \eta} h(\bm{\beta},\bm{\hat{X}})=\max\limits_{0 \leq \eta_{c_1}\leq \eta}h_a(\eta_{c_1},\bm{\beta}),$$ where $h_a(\eta_{c_1},\bm{\beta})=C_h(\|\bm{y}_1-\bm{X}_1\bm{\beta}\|_2+\eta_{c_1}\|\bm{\beta}\|_2)^2+D_h(\|\bm{y}_2-\bm{X}_2\bm{\beta}\|_2+\sqrt{\eta^2-\eta_{c_1}^2}\|\bm{\beta}\|_2)^2$;
	\item if $C_h< 0$, we have 
	$$\max \limits_{\|\bm{c}\bm{d}^T \|_F \leq \eta} h(\bm{\beta},\bm{\hat{X}})=\max\limits_{0 \leq \eta_{c_1}\leq \eta}h_b(\eta_{c_1},\bm{\beta}),$$ where
	\begin{eqnarray}
		&&\hspace{-10mm}h_b(\eta_{c_1},\bm{\beta})=
		\begin{cases}
			h_{b_1}(\eta_{c_1},\bm{\beta}), \text{ if }\| \bm{y}_1-\bm{X}_1 \bm{\beta}\|_2 \leq \eta \|\bm{\beta}\|_2,\\
			h_{b_2}(\eta_{c_1},\bm{\beta}), \text{ otherwise,}
		\end{cases} \nonumber
	\end{eqnarray}
	\begin{eqnarray}
		&&\hspace{-10mm}h_{b_1}(\eta_{c_1},\bm{\beta})=D_h(\|\bm{y}_2-\bm{X}_2\bm{\beta}\|_2+\sqrt{\eta^2-\eta_{c_1}^2}\|\bm{\beta}\|_2)^2, \nonumber\\
		&&\hspace{-10mm}h_{b_2}(\eta_{c_1},\bm{\beta})=C_h(\|\bm{y}_1-\bm{X}_1\bm{\beta}\|_2-\eta_{c_1}\|\bm{\beta}\|_2)^2 \nonumber\\
		&&\hspace{9mm}+ D_h(\|\bm{y}_2-\bm{X}_2\bm{\beta}\|_2+\sqrt{\eta^2-\eta_{c_1}}\|\bm{\beta}\|_2)^2.\nonumber
	\end{eqnarray}
	
\end{itemize}

\noindent\textbf{Transformation of the minimax problem}\\
After solving sub-problems above, the minimax problem \eqref{eq:minimax rank} can be transformed to a minimax problem for one vector and one scalar with a piece-wise max-type objective function. For example, 
if $D_g\geq 0$ and $C_h < 0$, \eqref{eq:minimax rank} can be represented as 
\begin{equation}
	\min_{\beta} \max_{0 \leq \eta_{c_1}\leq \eta} \max\{g_{a}(\eta_{c_1},\bm{\beta}),h_{b}(\eta_{c_1},\bm{\beta})\}. \label{eq:g_a, h_b}
\end{equation} 

Then we have the following two lemmas characterizing the nice properties of the sub-functions in the objective function. 
\begin{lem} \label{lm:weakly convex concave}
	If the norm of $\bm{\beta}$ is bounded, i.e. $\|\bm{\beta}\|_2 \leq B_{\beta}$, then we have
	\begin{enumerate}
		\item $g_a$ is weakly-concave in $\eta_{c_1}$ for any given $\bm{\beta}$ and weakly-convex in $\bm{\beta}$ for any given $\eta_{c_1}$;
		\item $h_b$ is a piece-wise function and each piece ($h_{b_1}$ or $h_{b_2}$) is weakly-concave in $\eta_{c_1}$ for any given $\bm{\beta}$ and weakly-convex in $\bm{\beta}$ for any given $\eta_{c_1}$.
	\end{enumerate}
\end{lem}
\begin{IEEEproof}
	Please refer to Appendix~\ref{app:weakly convex concave}.
\end{IEEEproof}
\begin{lem} \label{lm:unimodal}
	For any given $\bm{\beta}$, $g_a$, $g_{b_2}$, $h_{a}$ and $h_{b_2}$ are all unimodal functions with respect to $\eta_{c_1}$ that increase first and then decrease. 
\end{lem}
\begin{IEEEproof}
	Please refer to Appendix~\ref{app:unimodal}.
\end{IEEEproof}
Moreover, to deal with the piece-wise structure in the objective function, we further transform the minimax problem to several sub-problems. For example, \eqref{eq:g_a, h_b} can be transformed to three sub-problems:
\begin{enumerate}
	\item $\min \limits_{\beta} \max \limits_{0 \leq \eta_{c_1}\leq \eta} h_{b_1}(\eta_{c_1},\bm{\beta})$, \\
	s.t. $g_{a}(\eta_{c_1},\bm{\beta}) < h_{b_1}(\eta_{c_1},\bm{\beta}), \| \bm{y}_1-\bm{X}_1 \bm{\beta}\|_2 \leq \eta \|\bm{\beta}\|_2$;
	\item $\min \limits_{\beta} \max \limits_{0 \leq \eta_{c_1}\leq \eta} h_{b_2}(\eta_{c_1},\bm{\beta})$, \\
	s.t. $g_{a}(\eta_{c_1},\bm{\beta}) < h_{b_2}(\eta_{c_1},\bm{\beta}),\| \bm{y}_1-\bm{X}_1 \bm{\beta}\|_2 > \eta \|\bm{\beta}\|_2$.
	\item $\min \limits_{\beta} \max \limits_{0 \leq \eta_{c_1}\leq \eta} g_{a}(\eta_{c_1},\bm{\beta})$, s.t. $g_{a}(\eta_{c_1},\bm{\beta}) \geq h_{b}(\eta_{c_1},\bm{\beta})$;
\end{enumerate}

For the sub-problem 1), the maximization on $\eta_{c_1}$ can be solved exactly and the saddle-point can be easily derived.

For sub-problems 2) and 3), we will ignore the constraints first and derive the saddle-point of the minimax problem, and then check the constraints. For example, for sub-problem 2), we assume that $\|\bm{\beta}\|_2 \leq B_{\beta}$, which is reasonable in reality, and have that: 
\begin{itemize}
	\item the feasible set $\{\bm{\beta}:\|\bm{\beta}\|_2 \leq B_{\beta}\} \times [0,\eta]$ is convex and compact;
	\item the objective function is weakly-convex-weakly-concave by Lemma~\ref{lm:weakly convex concave};
	\item the saddle-point exists by Lemma~\ref{lm:unimodal}. 
\end{itemize}
Based on those properties, we are able to apply a first-order algorithms proposed by \cite{liu2021first} to solve the non-convex non-concave minimax problem as in sub-problem 2) and derive the nearly $\epsilon$-stationary solution. In particular, define $\mathcal{Z}= \{\bm{\beta}:\|\bm{\beta}\|_2 \leq B_{\beta}\} \times [0,\eta]$ and the mapping $H(\bm{z}) \coloneqq (\partial_{\bm{\beta}}h_{b_2}(\eta_{c_1},\bm{\beta}),\partial_{\eta_{c_1}}[-h_{b_2}(\eta_{c_1},\bm{\beta})])^T$, where $\bm{z}=(\bm{\beta},\eta_{c_1})$. The minty variational inequality (MVI) problem corresponding to the saddle-point problem in sub-problem 2) is to find $\bm{z}^* \in \mathcal{Z}$ such that $\langle \bm{\xi},\bm{z}-\bm{z}^*\rangle \geq 0, \forall \bm{z} \in \mathcal{Z}, \forall \bm{\xi} \in H(\bm{z})$. Then the saddle-point problem can be solved through the lens of MVI. In \cite{liu2021first}, the proposed inexact proximal point method consists of approximately solving a sequence of strongly monotone MVIs constructed by adding a strongly monotone mapping to $H(\bm{z})$ with a sequentially updated proximal center. Thus, the complex non-convex non-concave minmax problem can be decomposed into a sequence of easier strongly-convex strongly-concave problems.

\section{Numerical Results}\label{sec:numerical}
	In this section, we provide numerical examples to illustrate the results in this paper. We conduct experiments on a synthetic dataset and two real-world datasets: 

\noindent 1. \textbf{Synthetic Dataset (SD)}: it contains 200 rows for two groups with 5 features. We suppose that the numbers of samples in two groups are the same, i.e. $m=n-m=100$. For two different groups, the samples are generated by
\begin{eqnarray}
	\bm{y}_1=\bm{X}_1\bm{\beta}_{0,1}+\bm{c}_1+\bm{\epsilon}, ~
	\bm{y}_2=\bm{X}_2\bm{\beta}_{0,2}+\bm{\epsilon}, \label{eq:construct eq}
\end{eqnarray} 
where elements in $\bm{X}_1$ and $\bm{X}_2$ are uniformly distributed on $(0,10)$, $\bm{\beta}_{0,1}=[1,1,1,1,1]^T$, $\bm{c}_1=[1,\cdots,1]^T$, $\bm{\beta}_{0,2}=[1.1,1.1,1.1,1.1,1.1]^T$ and noise $\bm{\epsilon} \sim \mathcal{N}(0,1)$. Under this setup, we verify the assumption in Propositions~\ref{prop:case 2} and have that $\eta^2 \geq \eta_{min}^2= \max\left\{\frac{(n+1)v_{X_1,p}}{m(m+1)},\frac{(n+1)v_{X_2,p}}{(n-m+1)(n-m)}\right\}=15.98^2$ while the mean energy of a sample is $\eta_D=29.08$, which indicates that the assumption on $\eta$ is reasonable. 

\noindent 2. \textbf{Law School Dataset (LSD)} \cite{wightman1998lsac}: it contains 1,823 records for law students who took the bar passage study for law school admission, with gender as the sensitive attribute and undergraduate GPA as the target variable. The dimension of features is 8. There are 999 samples and 824 samples for two genders respectively. For the assumption on $\eta$, we have $\eta \geq \eta_{min}=2.44$ and $\eta_D=2.86$. 

\noindent 3. \textbf{Medical Insurance Cost Dataset (MICD)} \cite{lantz2019machine}: it contains 1,338 medical expense examples for patients in the United States. In our experiment, we use gender as the sensitive attribute, charged medical expenses as the target variable, and consider $5$ features. There are 662 samples and 676 samples for two genders respectively. Then we verify the assumption on $\eta$ and have that $\eta \geq \eta_{min}=1.58$ with $\eta_D=2.34$.

For comparison purpose, we will introduce an unrobust fair regression model that does not consider the existence of the adversary and minimizes the objective function with respect to the original dataset $\{\bm{X},\bm{y},\bm{G}\}$. In particular, the unrobust fair model is
\begin{eqnarray}
	\bm{\beta}_{fair}=\arg \min_{\bm{\beta}} f(\bm{\beta},\bm{X},\bm{y},\bm{G})+ \lambda F(\bm{\beta},\bm{X},\bm{y},\bm{G}). \nonumber
\end{eqnarray}
Moreover, for the rank-one attack scheme, we also compare our proposed adversarially robust model with other fair regression models, including the fair linear regression (FLR) model and fair kernel learning (FKL) model \cite{perez2017fair}. The optimal regression coefficient for each model is derived by fitting the model on the original dataset $\{\bm{X},\bm{y},\bm{G}\}$. To obtain the performance of each model on the poisoned dataset, we apply the derived optimal regression coefficient on the poisoned dataset, $\{\hat{\bm{X}},\hat{\bm{y}},\hat{\bm{G}}\}$, and calculate the MSE as well as the group fairness gap. 

\subsection{Attack with one adversarial data point}
\begin{figure}[h]
	\centering
	\subfigure[Error Gap v.s. $R^2$, $\lambda=0.2$]{
		\label{fig:Gap_R2 0.2}
		\includegraphics[width=0.22\textwidth]{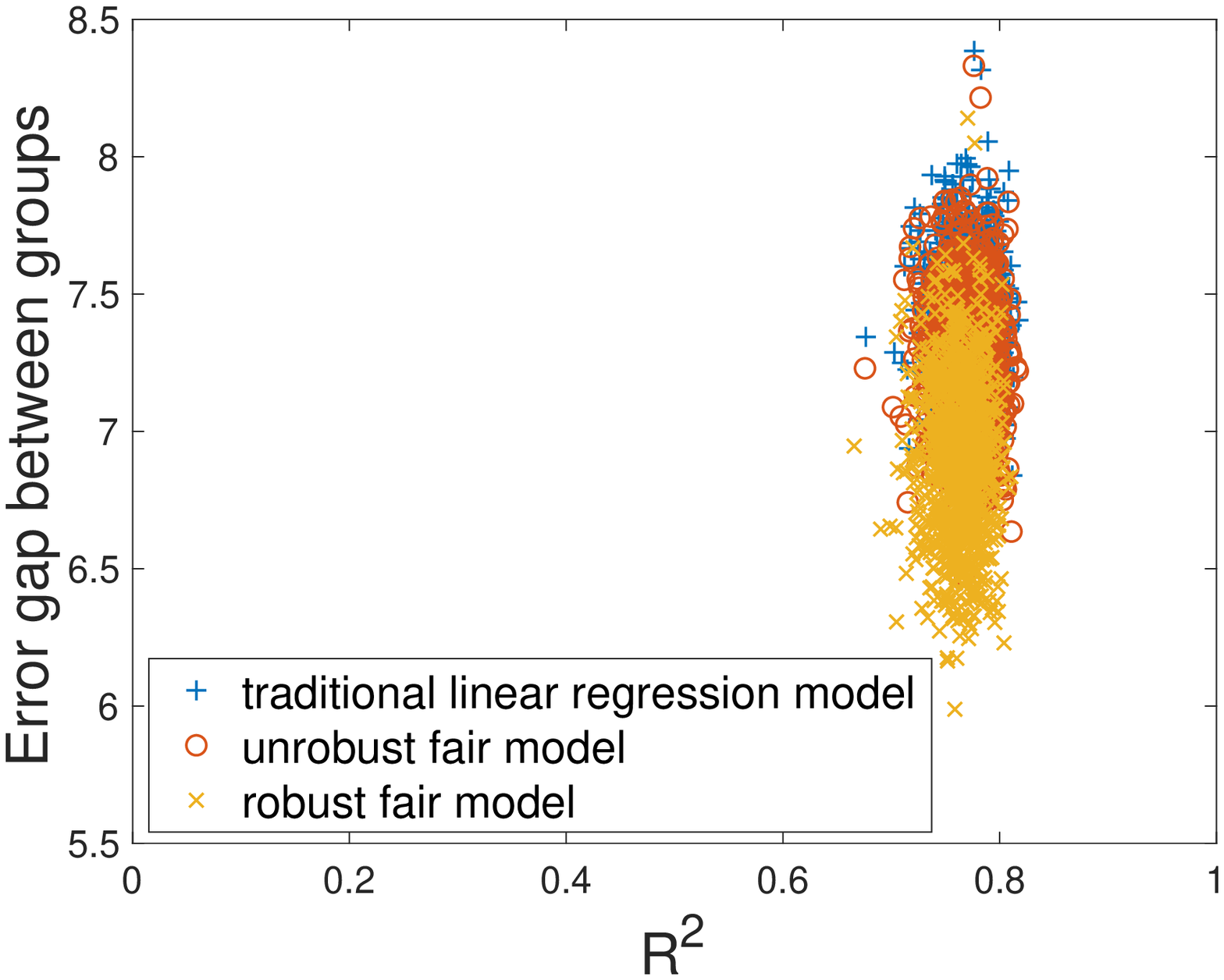}
	}
	\subfigure[Error Gap v.s. $R^2$, $\lambda=0.8$]{
		\label{fig:Gap_R2 0.8}
		\includegraphics[width=0.22\textwidth]{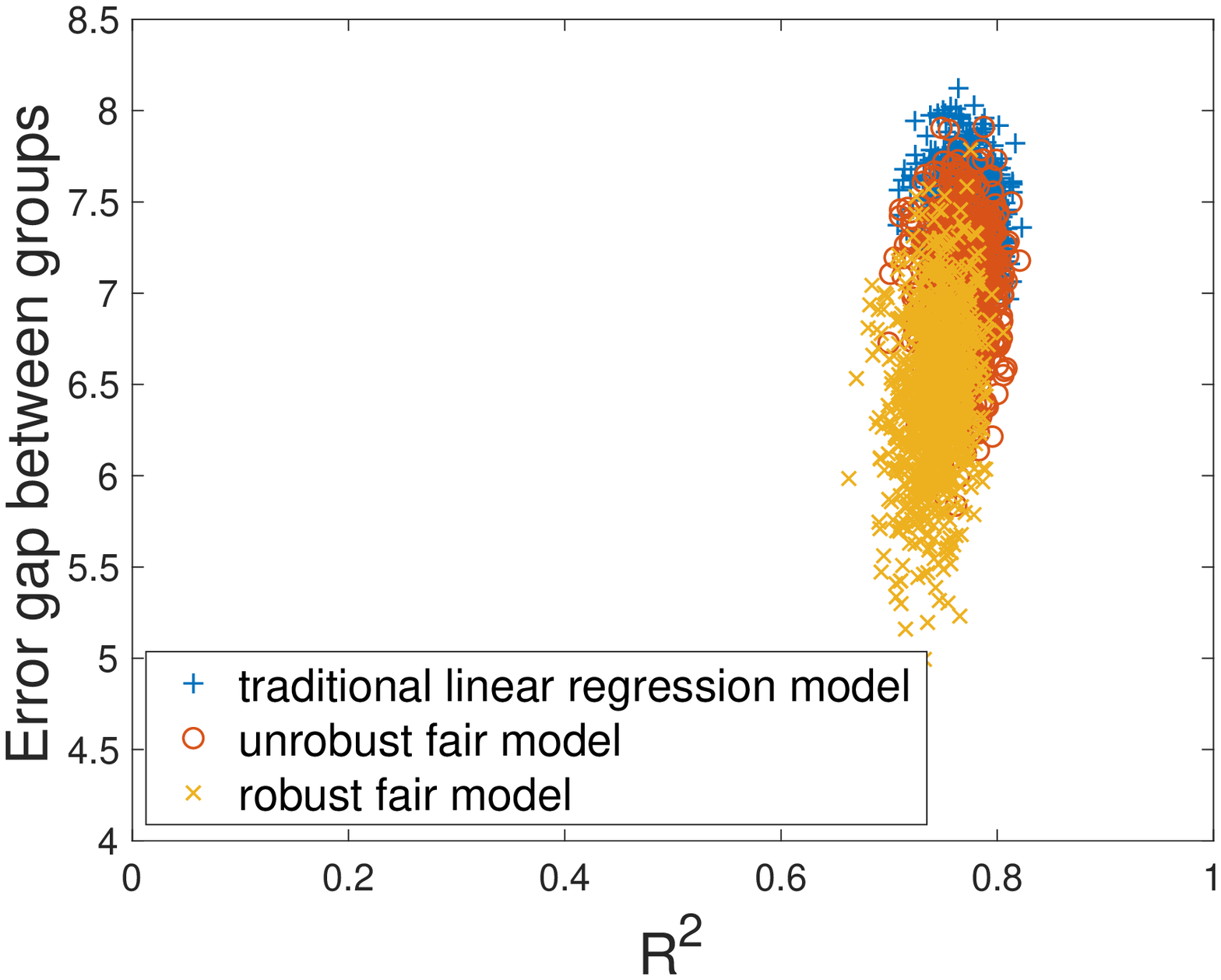}
	}
	\caption{SD: comparison of robust fair model, unrobust fair model and traditional linear model (attack with one adversarial data point).}
\end{figure}

Firstly, for SD, by choosing $\eta=\eta_D$, we explore the performance differences among the proposed robust fairness-aware model, unrobust fair model and traditional linear model (ordinary linear regression model). In Fig.~\ref{fig:Gap_R2 0.2} and Fig.~\ref{fig:Gap_R2 0.8}, following \eqref{eq:construct eq}, we construct 500 datasets relying on the randomness in $\bm{\epsilon}$. For $\lambda=0.2< \min\{\frac{m+1}{n+1},\frac{n-m+1}{n+1}\}$ (which implies $C_{h_1}\geq 0, D_{g_2} \geq 0$), according to Theorem~\ref{thm:L upper-bound}, the best adversarial point is $\tilde{\bm{x}}_0=\eta \frac{\bm{b}}{\|\bm{b}\|_2}$.
As shown in Fig.~\ref{fig:Gap_R2 0.2}, the group fairness gap for the proposed robust fairness-aware model is smaller than that of the unrobust fair model, while the measure of goodness of fit $R^2$ remains similar. In the meantime, since $\bm{\beta}_{fair}$ has taken the fairness issue into consideration, its performance is better than the traditional linear regression model. Likewise, for $\lambda=0.8> \max\{\frac{m+1}{n+1},\frac{n-m+1}{n+1}\}$ (which implies $C_{h_1}< 0, D_{g_2} < 0$), according to Theorem~\ref{thm:L upper-bound}, the best adversarial point will be in the form $\tilde{\bm{x}}_0 \perp \bm{b}$ or $\tilde{\bm{x}}_0=\eta \frac{\bm{b}}{\|\bm{b}\|_2}$ based on the value of $g_i(\bm{\beta}^*_{rob})$ and $h_i(\bm{\beta}^*_{rob}), i=1,2$. As shown in Fig.~\ref{fig:Gap_R2 0.8}, the performance results are similar to the case $\lambda=0.2$.

\begin{figure}
	\centering
	\subfigure[LSD: MSE v.s. $\lambda$]{
		\label{fig:LSD: MSE1}
		\includegraphics[width=0.22\textwidth]{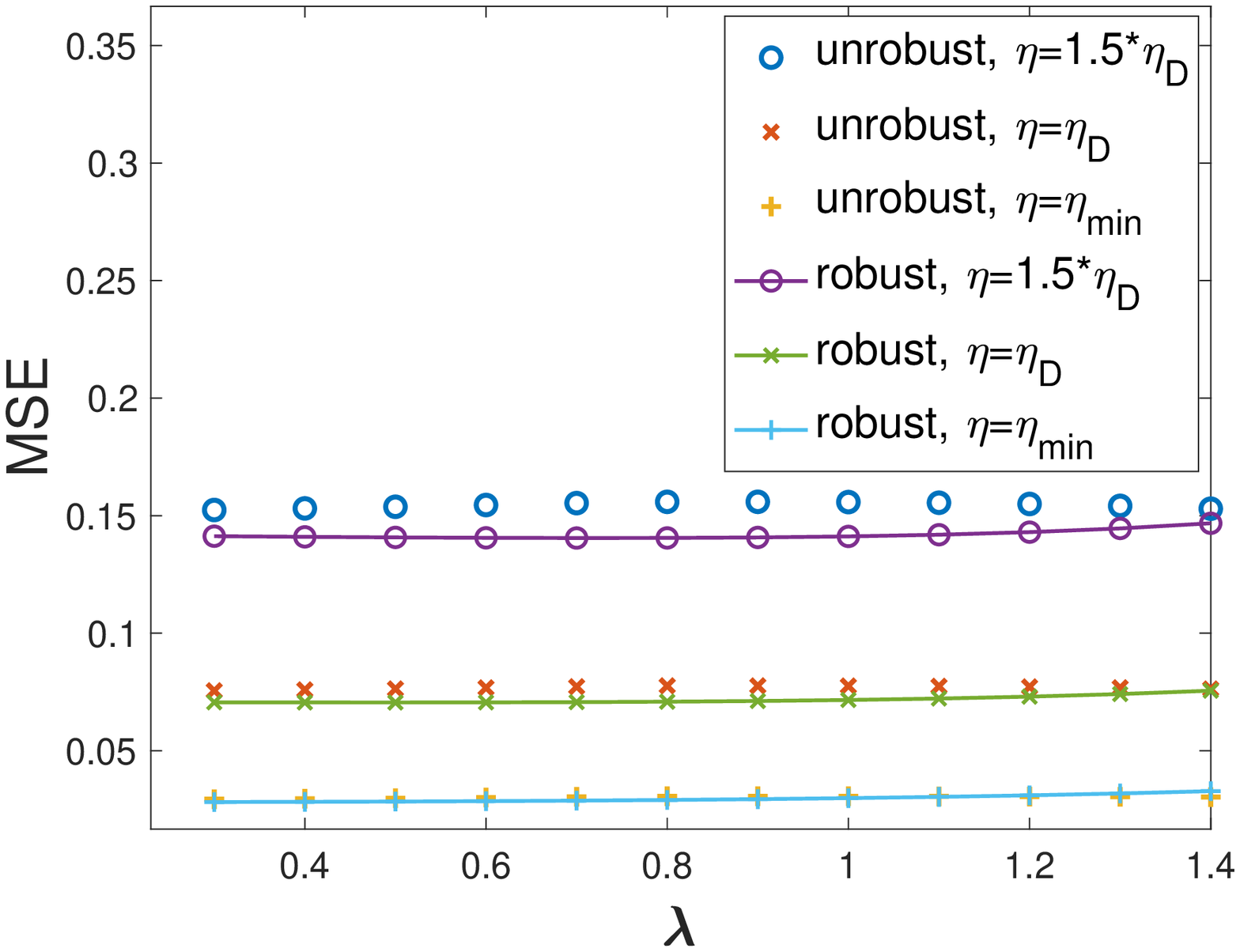}
	}
	\subfigure[LSD: Group fairness gap v.s. $\lambda$]{
		\label{fig:LSD: Error gap1}
		\includegraphics[width=0.22\textwidth]{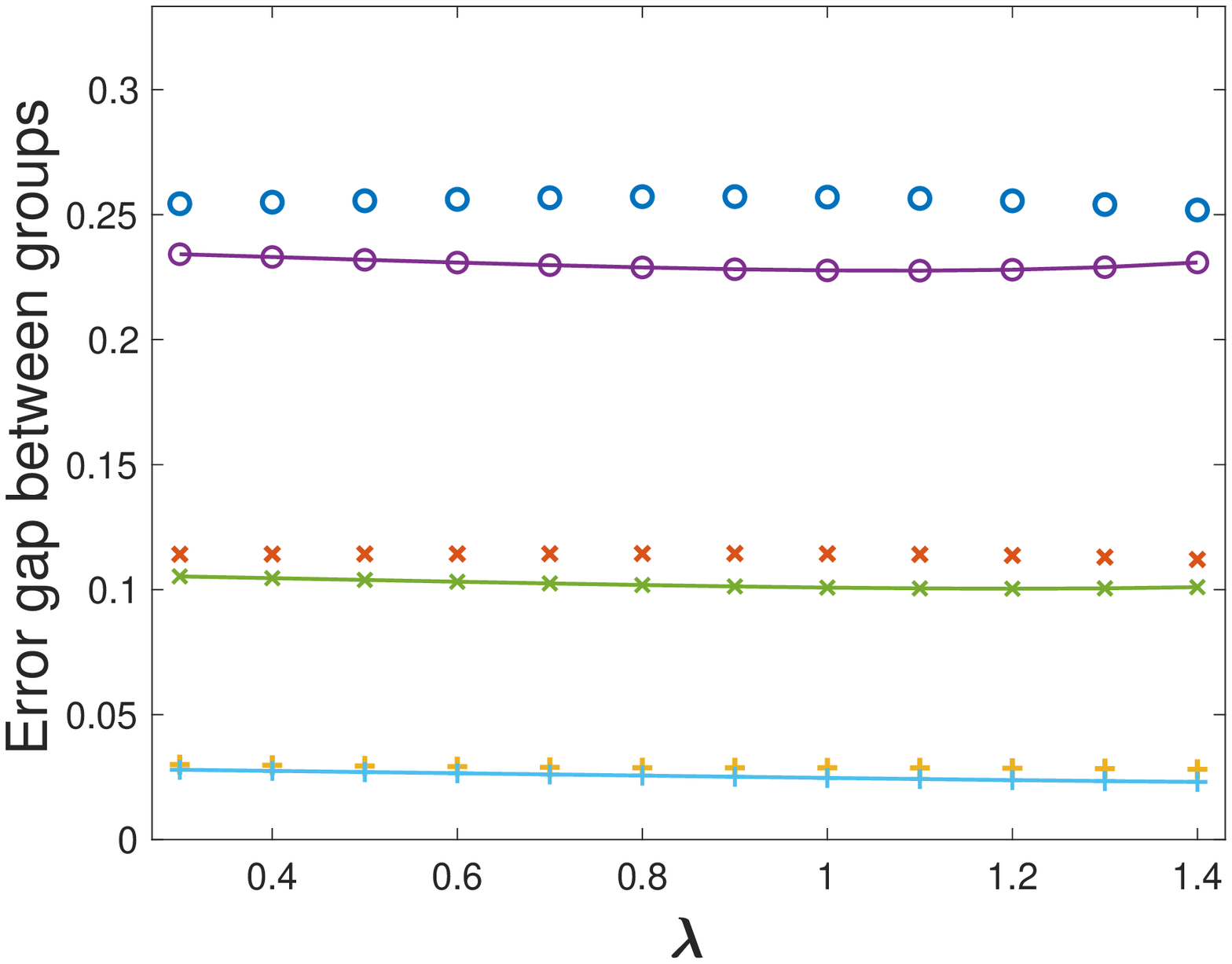}
	}
	\subfigure[MICD: MSE v.s. $\lambda$]{
		\label{fig:MICD: MSE1}
		\includegraphics[width=0.22\textwidth]{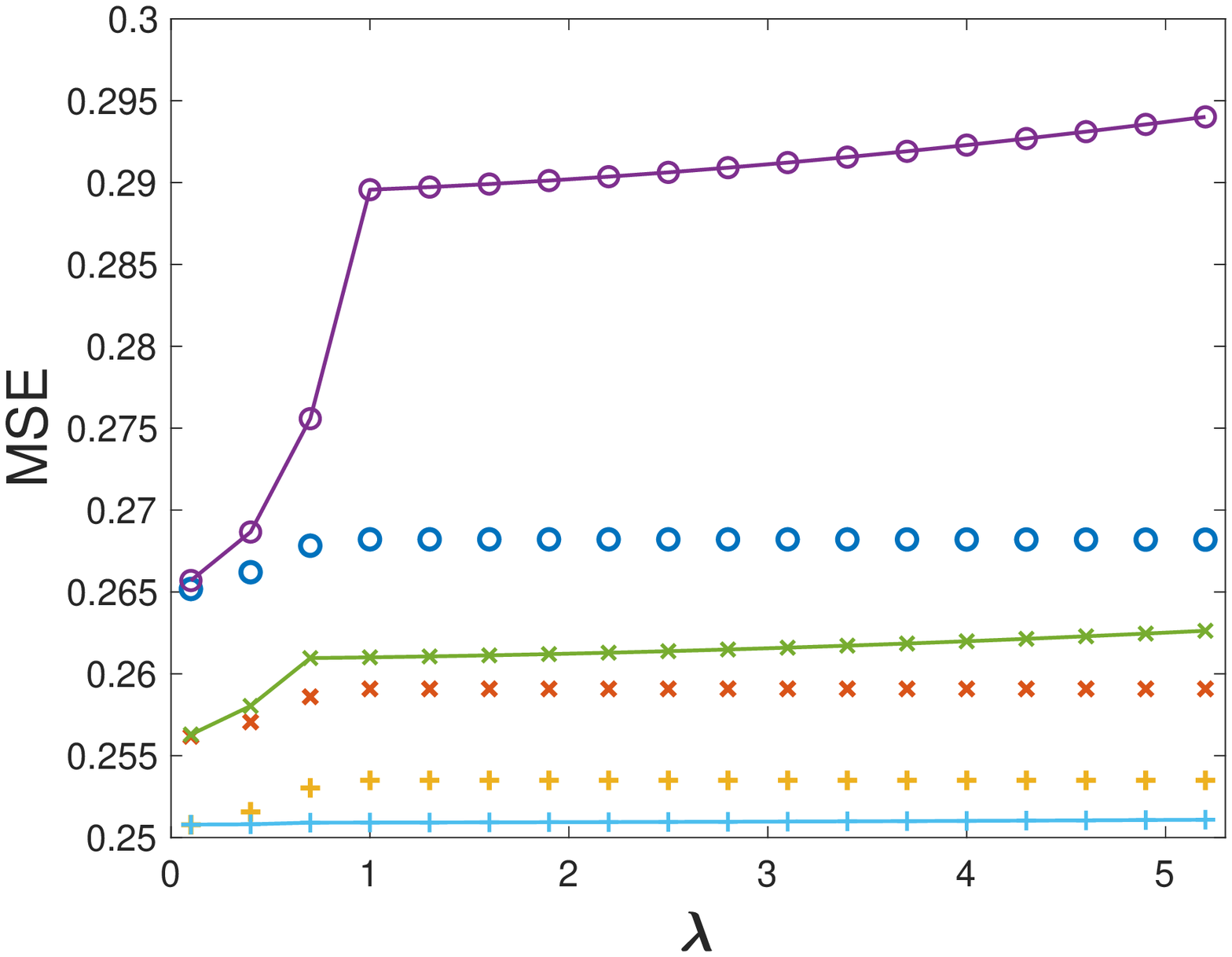}
	}
	\subfigure[MICD: Group fairness gap v.s. $\lambda$]{
		\label{fig:MICD: Error gap1}
		\includegraphics[width=0.22\textwidth]{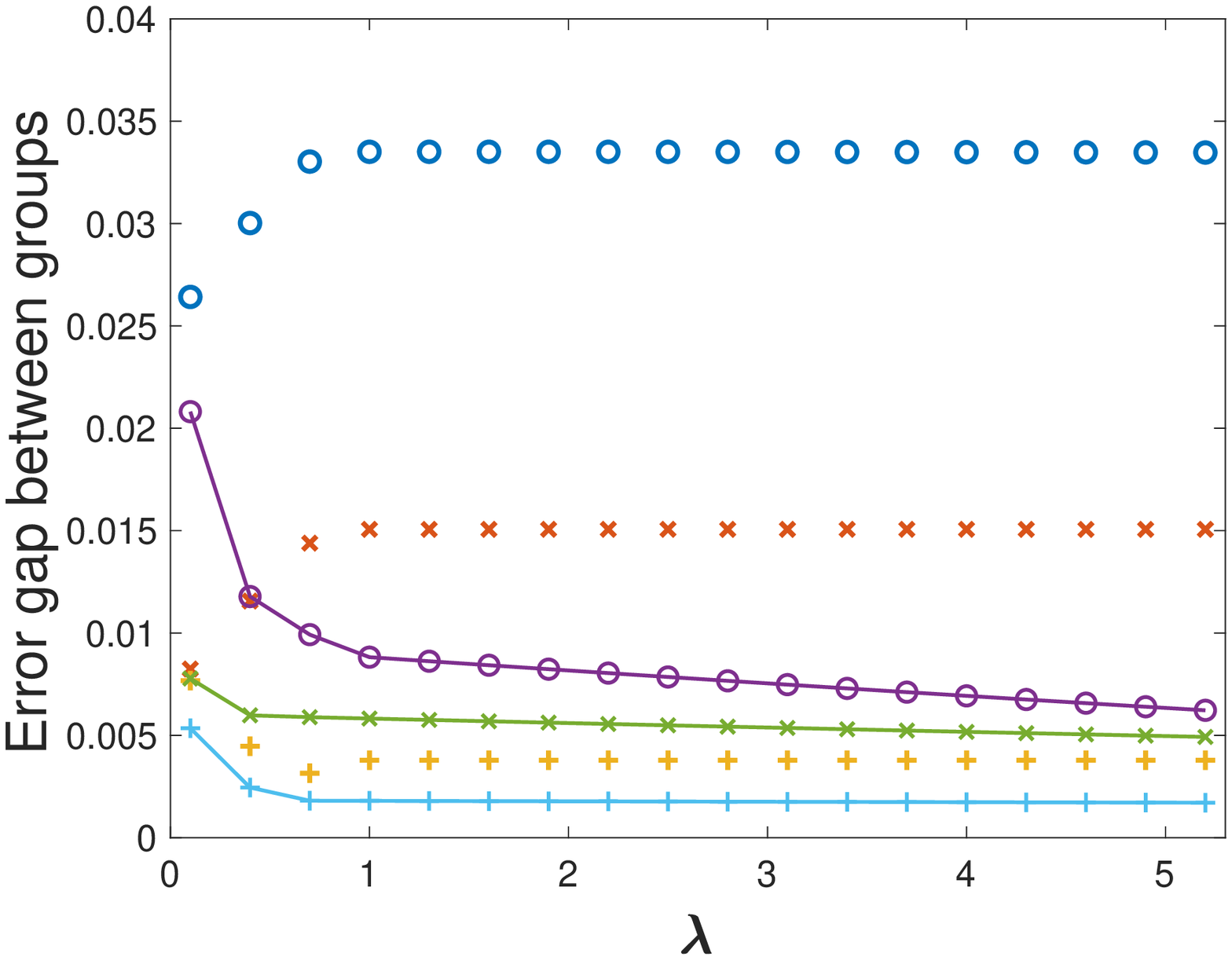}
	}
	\caption{Effects of $\lambda$ and $\eta$ on MSE and the group fairness gap (attack with one adversarial data point).}
	\label{fig:eta1}
\end{figure}

Secondly, we explore the effects of the energy constraint parameter $\eta$ as well as the trade-off parameter $\lambda$ on two real-world datasets, LSD and MICD. We have three energy levels, $\eta=\eta_{min}$,  $\eta=\eta_D$ and $\eta=1.5\eta_D$. As shown in Fig.~\ref{fig:eta1}, when $\eta$ is small, under different choices of $\lambda$, MSE and the group fairness gap for the robust fairness-aware model are both smaller than those for the unrobust fair model, which indicates that the proposed model has better robustness and achieves better performance in both accuracy and fairness. However, for MICD, when $\eta=1.5\eta_D$, the MSE for the robust fair model becomes larger than that of the unrobust model as the power of the adversarial data point is large, which in turn affects the prediction performance considerably.

\subsection{Rank-one attack}
\begin{figure}[h]
	\centering
	\subfigure[LSD: MSE v.s. $\lambda$]{
		\label{fig:LSD: MSE}
		\includegraphics[width=0.22\textwidth]{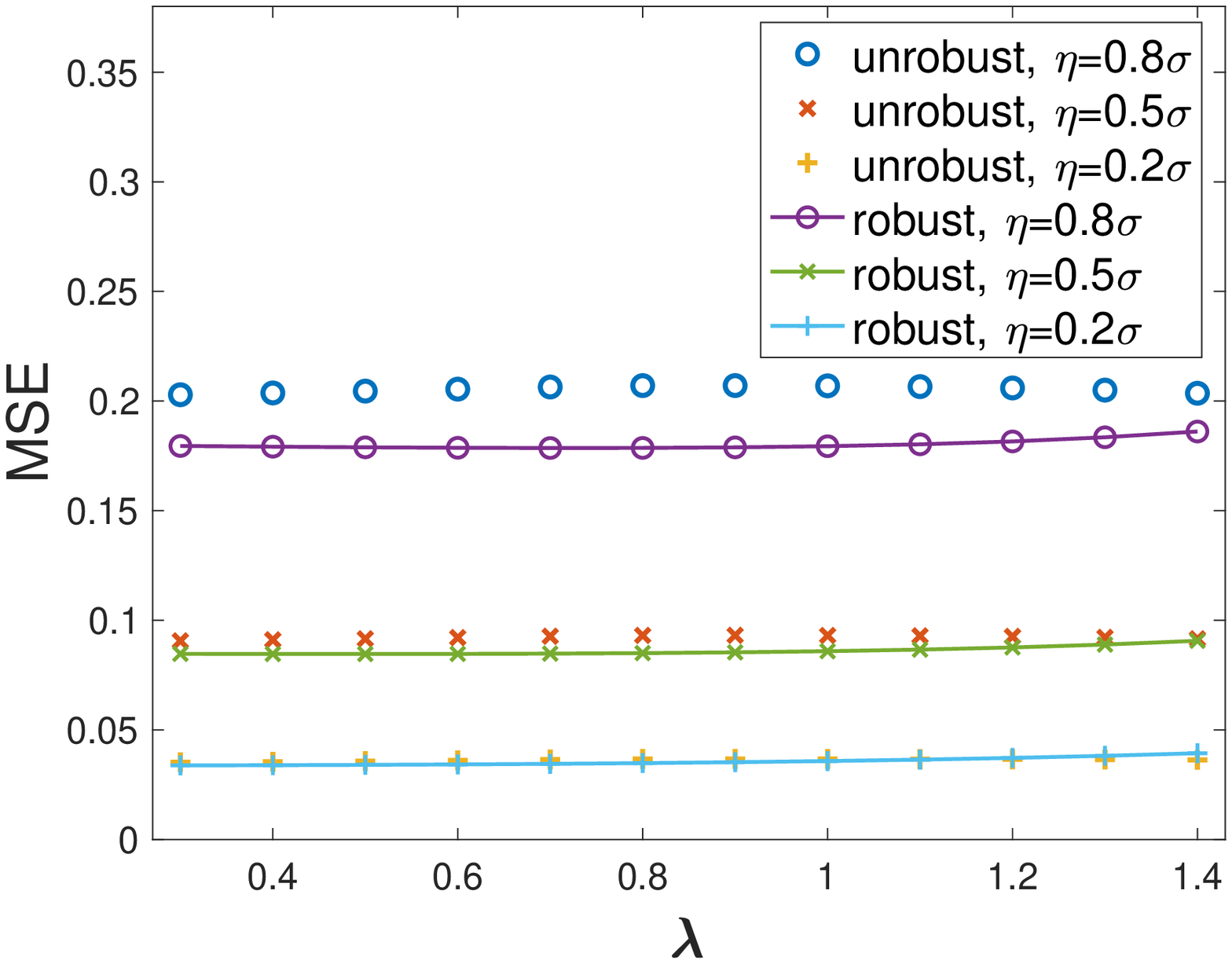}
	}
\subfigure[LSD: Group fairness gap v.s. $\lambda$]{
	\label{fig:LSD: Error gap}
	\includegraphics[width=0.22\textwidth]{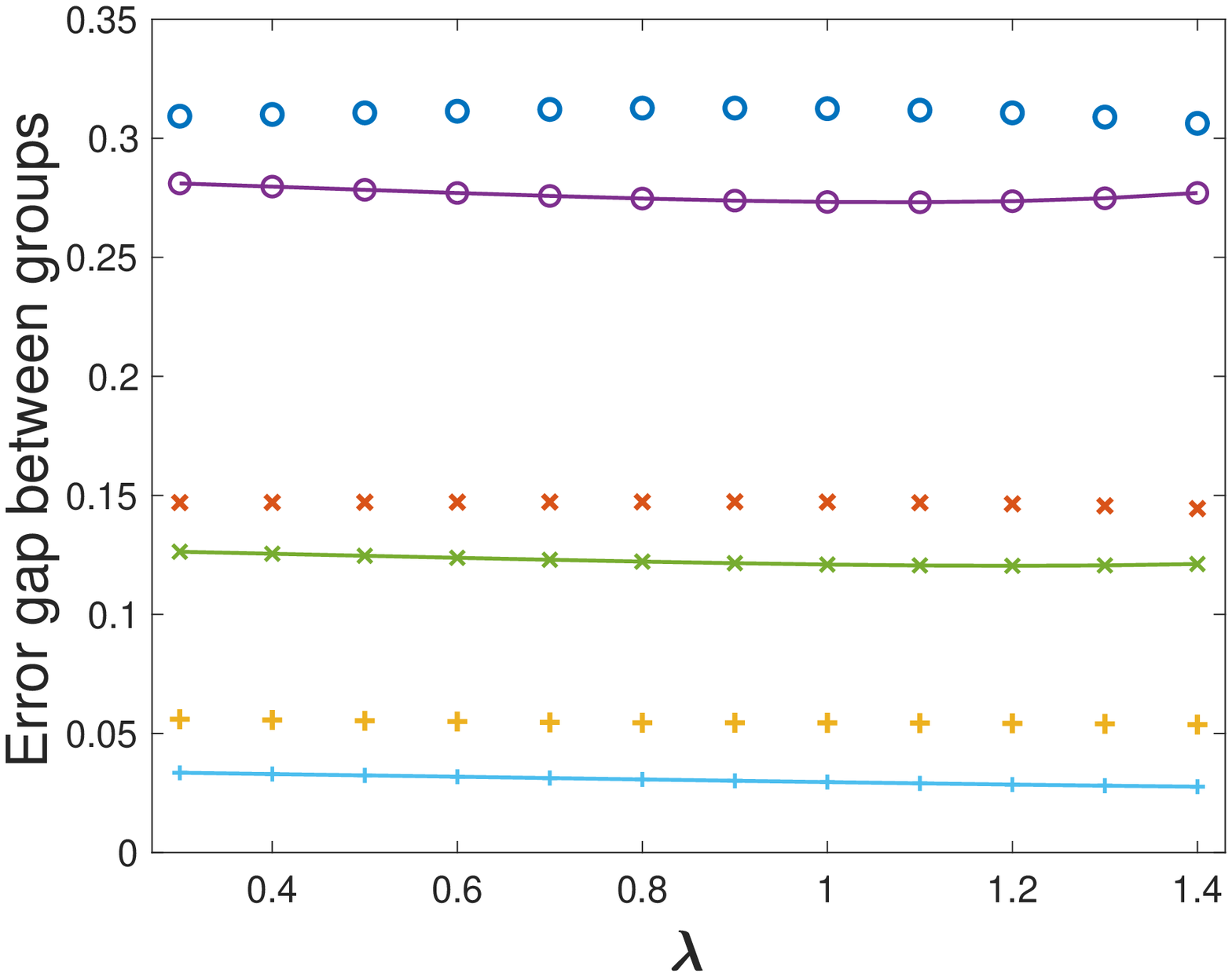}
}
	\qquad
	\subfigure[MICD: MSE v.s. $\lambda$]{
		\label{fig:MICD: MSE}
		\includegraphics[width=0.22\textwidth]{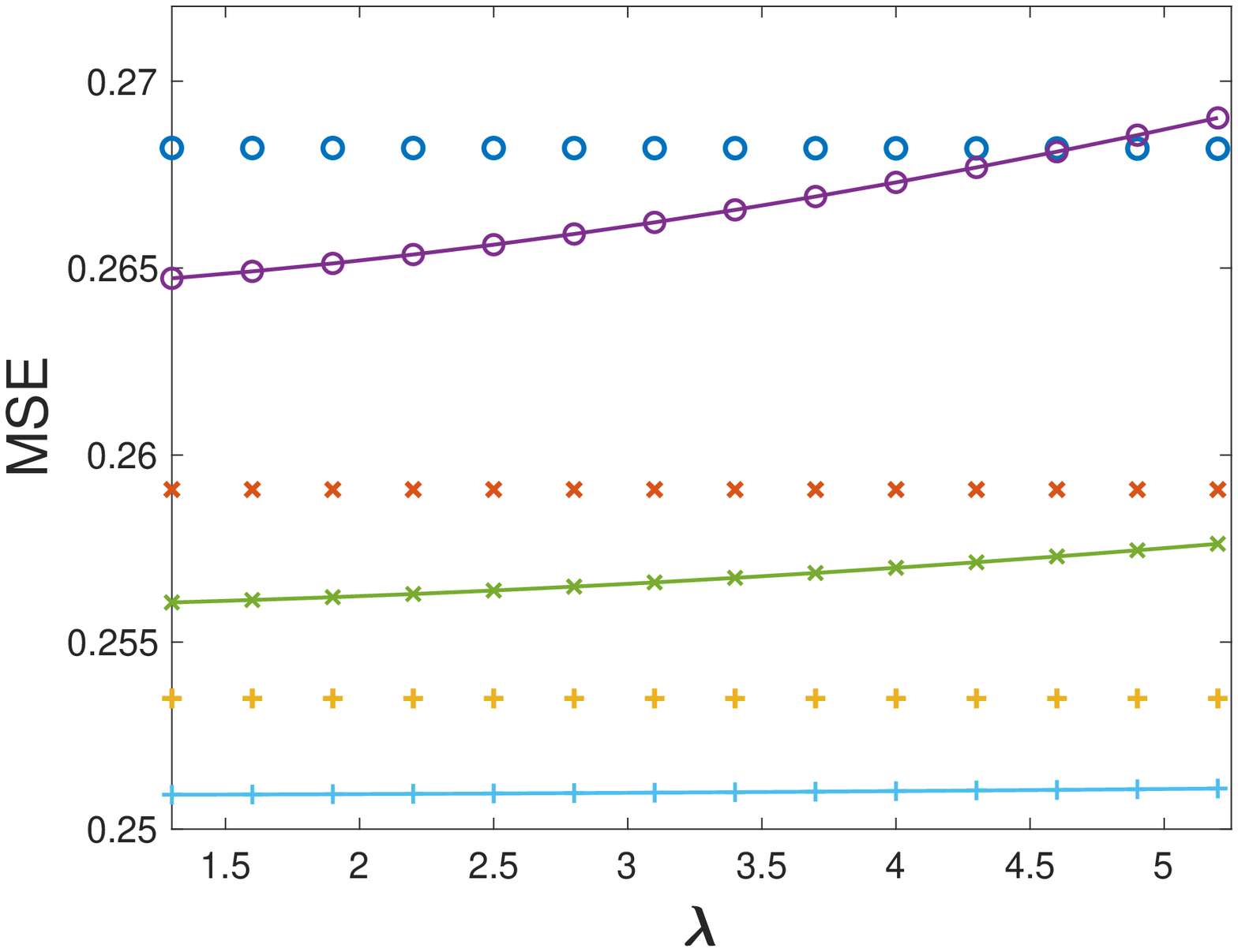}
	}
	\subfigure[MICD: Group fairness gap v.s. $\lambda$]{
		\label{fig:MICD: Error gap}
		\includegraphics[width=0.22\textwidth]{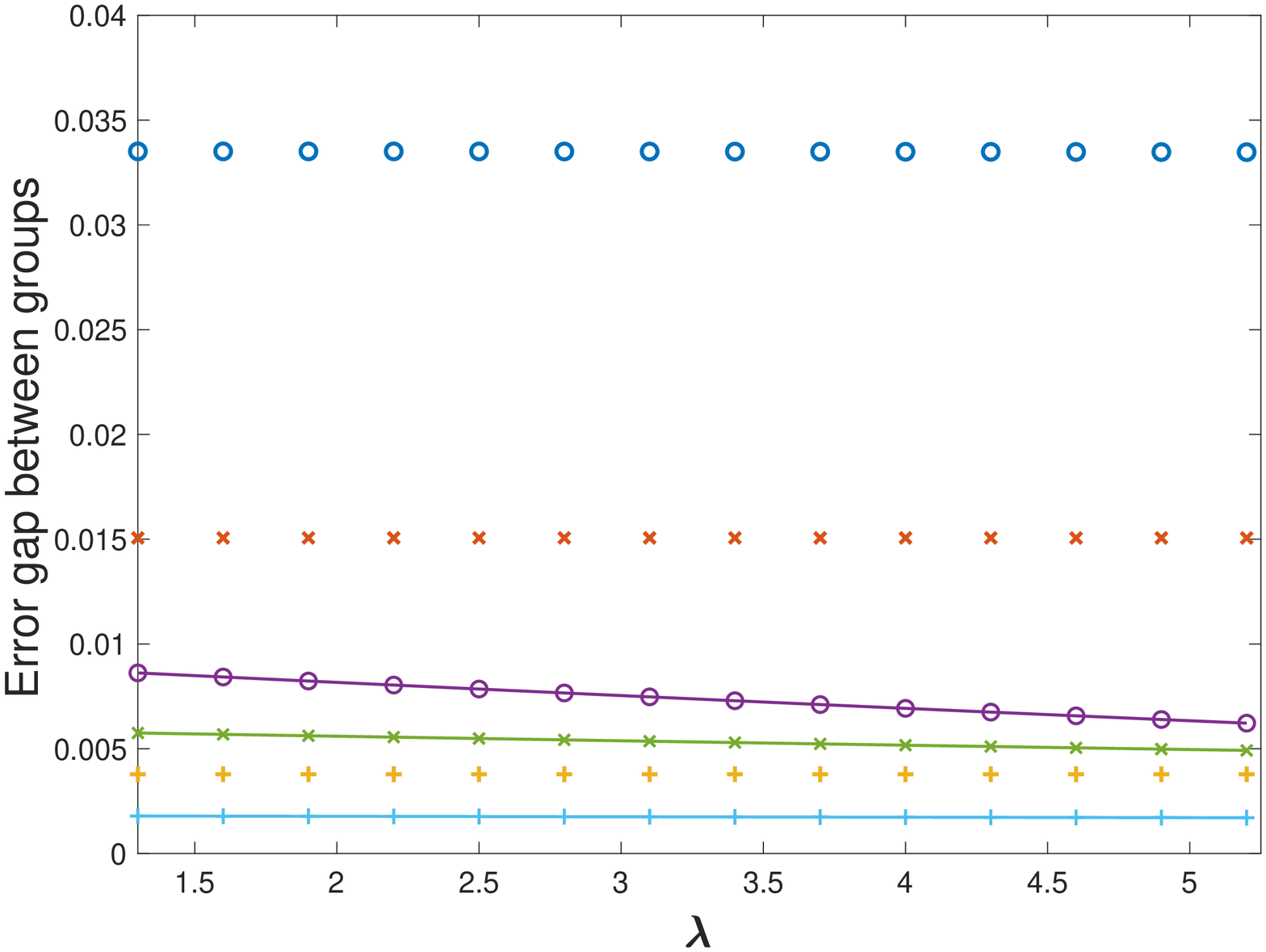}
	}
	\caption{Effects of $\lambda$ and $\eta$ on MSE and fairness gap (rank-one attack)}
	\label{fig:eta}
\end{figure}

In the first experiment, we explore the effects of the energy constraint parameter $\eta$ as well as the trade-off parameter $\lambda$. We carry out the attack with three different energy levels, $\eta=0.2\sigma$,  $\eta=0.5\sigma$ and $\eta=0.8\sigma$, where $\sigma$ is the smallest singular value of the feature matrix of the training data. As shown in Fig.~\ref{fig:eta}, we first observe that MSE and the group fairness gap for the adversarially robust model are almost always smaller than those for the unrobust fair model, which illustrates that the proposed robust model achieves better performance in both accuracy and fairness. We also notice that the performance of the adversarially robust model differs under different choices of $\lambda$. In particular, as $\lambda$ increases, the value of MSE also increases because we care more about fairness and give more weight to the fairness-related term in the objective function. 
Especially, as shown in Fig.~\ref{fig:MICD: MSE}, when the energy constraint is comparable to the smallest singular value of the feature matrix ($\eta=0.8\sigma$) and the trade-off parameter $\lambda$ is large ($\lambda=5.2$), the MSE of the robust model becomes larger than that of the unrobust model as the limitation on the adversary is small, which in turn affects the prediction performance considerably.

\begin{figure}[h]
	\centering
	\subfigure[Original dataset]{
		\label{fig:ori}
		\includegraphics[width=0.22\textwidth]{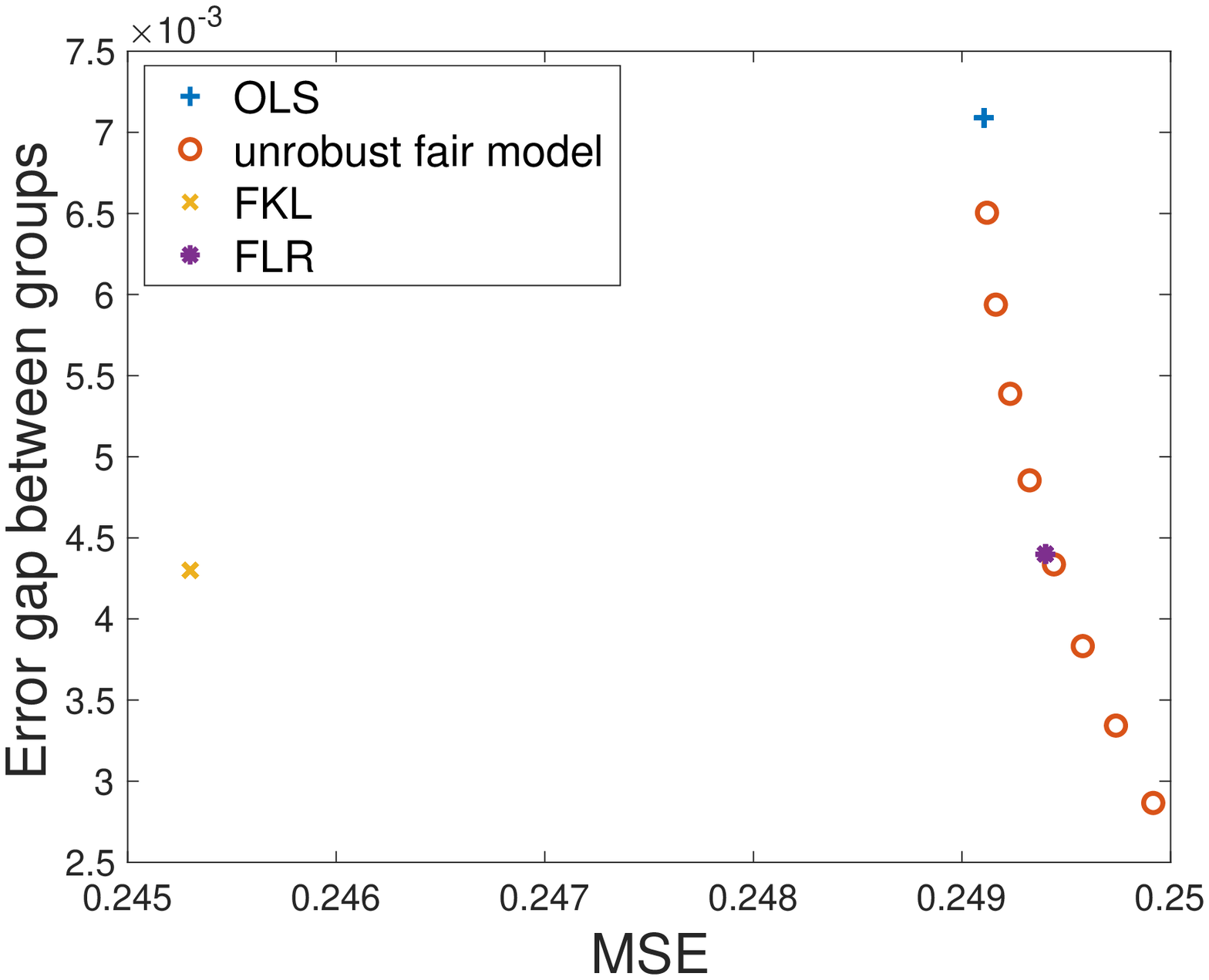}
	}
	\subfigure[Poisoned dataset]{
		\label{fig:poisoned}
		\includegraphics[width=0.22\textwidth]{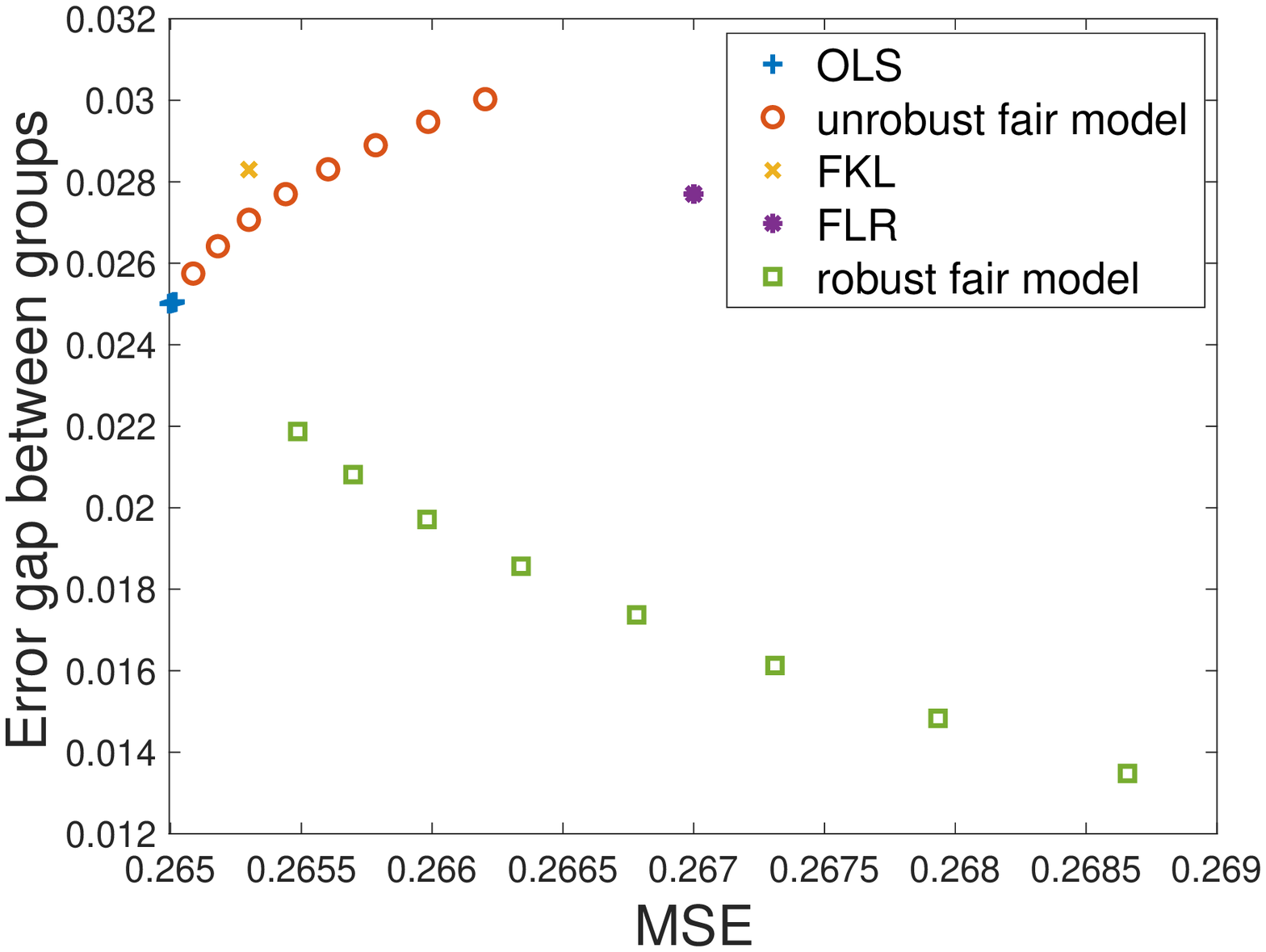}
	}
	\caption{MICD: Group fairness gap v.s. MSE (rank-one attack).}
	\label{fig:compare}
\end{figure}

In the second experiment, we compare our proposed adversarially robust fair model with other fair regression models. In Fig.~\ref{fig:compare}, we provide the performance of different regression models on the original dataset as well as the poisoned dataset with $\eta=0.5\sigma$. For the unrobust fair model and adversarially robust fair model, since the choice of the trade-off parameter $\lambda$ will affect the model performance, we explore models with various choices of $\lambda$. As shown in Fig.~\ref{fig:ori}, on the original dataset, the overall performance of FKL is better than other models, since it is a nonlinear model based on kernels. FLR has similar performance with the proposed unrobust fair regression model (with certain choice of $\lambda$). Moreover, for the unrobust fair model, it is observed that as $\lambda$ increases, the group fairness gap decreases while the MSE increases. However, on the poisoned dataset, as shown in Fig.~\ref{fig:poisoned}, the performance of FKL and FLR has been severely impacted. In particular, for FKL (which is the optimal model on the original dataset), the value of the group fairness gap has been increased from $4.3 \times 10^{-3}$ to $2.8 \times 10^{-2}$, and the value of MSE also increases. Similar observations can be found for FLR. Besides, for the unrobust fair model, we observe a concave curve in the group fairness gap v.s. MSE plot, which is convex in the original dataset. Thus, we conclude that fair regression models are vulnerable to adversarial attacks and may not preserve their performance in adversarial environment. On the contrary, for the adversarially robust model, the curve between the group fairness gap and MSE locates in the lower left corner and is convex. Thus, by appropriately choosing the value of $\lambda$, a model that performs well in terms of both fairness and prediction accuracy can be obtained. 

\section{Conclusion}  \label{sec:conclude}
In this paper, we have proposed a minimax framework to characterize the best attacker that generates the optimal poisoned point or rank-one attack for the original dataset, as well as the adversarially robust fair defender that can achieve the best performance in terms of both prediction accuracy and fairness guarantee, in the presence of the best attacker. We have discussed two types of attack schemes and provided the corresponding methods to solve the proposed nonsmooth nonconvex-nonconcave minimax problems. Moreover, we have performed numerical experiments on synthetic data and two real-world datasets, and shown that the proposed adversarially robust fair models can achieve better performance in both prediction accuracy and fairness guarantee than other fair regression models with a proper choice of $\lambda$.

\appendices
\section{Proof of Theorem~\ref{thm:L upper-bound}} \label{app:L upper-bound}
We will prove the maximum value of $L$ under two cases: $G_0=1$ and $G_0=2$ separately. For $G_0=1$, we will show  
$
\max \limits_{\substack{(\bm{x}_0,y_0,1),\\\text{s.t. } \|[\bm{x}_0^T,y_0]\|_2\leq\eta }} L_1 \overset{(b)}{=} \max\{g_1(\bm{\beta}),h_1(\bm{\beta})\}.
$
Similarly, for $G_0=2$, we will have that
$
\max \limits_{\substack{(\bm{x}_0,y_0,2),\\\text{s.t. } \|[\bm{x}_0^T,y_0]\|_2\leq\eta }} L_2 \overset{(c)}{=} \max\{g_2(\bm{\beta}),h_2(\bm{\beta})\}.
$
Then $(a)$ follows directly from $(b)$ and $(c)$. Since the case $G_0=2$ is similar to the case $G_0=1$, we will only verify the equality $(b)$.

Firstly, for the adversarial point, under the constraint that $\|\tilde{\bm{x}}_0^T\|_2=\|[\bm{x}_0^T,y_0]\|_2\leq\eta,$
we have
\begin{eqnarray}
	0 \leq \|\tilde{\bm{x}}_0^T \bm{b}\|_2^2 \leq \eta^2 \|\bm{b}\|_2^2 = \eta^2(1+\|\bm{\beta}\|_2^2). \label{eq:energey}
\end{eqnarray}
Then we notice that
\begin{eqnarray}
	&& \hspace{-10mm}L_1\overset{(d)}{\leq} \max\left\{\left(\frac{\lambda}{m+1}+\frac{1}{n+1}\right)\eta^2(1+\|\bm{\beta}\|_2^2) \right. \nonumber\\
	&& \left. +\left(\frac{\lambda}{m+1}+\frac{1}{n+1}\right)\|\bm{y_1}-\bm{X}_1 \bm{\beta}\|_2^2 \right. \nonumber\\
	&& \left.  +\left(-\frac{\lambda}{n-m}+\frac{1}{n+1}\right)\|\bm{y_2}-\bm{X}_2 \bm{\beta}\|_2^2,\right. \nonumber\\
	&& \left. \max\{0,-\frac{\lambda}{m+1}+\frac{1}{n+1}\}\eta^2(1+\|\bm{\beta}\|_2^2)   \right. \nonumber\\
	&& \left.  +\left(-\frac{\lambda}{m+1}+\frac{1}{n+1}\right)\|\bm{y_1}-\bm{X}_1 \bm{\beta}\|_2^2\right.\nonumber\\
	&& \left.  +\left(\frac{\lambda}{n-m} +\frac{1}{n+1}\right)\|\bm{y_2}-\bm{X}_2 \bm{\beta}\|_2^2\right\} \nonumber\\
	&=& \max\{g_1(\bm{\beta}), h_1(\bm{\beta})\},\nonumber
\end{eqnarray}
where $(d)$ is from \eqref{eq:energey}. Then we verify the achievability of the equality in $(d)$. Define a set $B_1 \coloneqq\{\bm{\beta}: g_1(\bm{\beta}) \geq h_1(\bm{\beta})$ $ =\{\bm{\beta}:\frac{1}{m+1}\|\bm{y_1}-\bm{X}_1 \bm{\beta}\|_2^2 -\frac{1}{n-m}\|\bm{y_2}-\bm{X}_2 \bm{\beta}\|_2^2 \geq \max\left\{-\frac{1}{2(m+1)}-\frac{1}{2\lambda(n+1)},-\frac{1}{m+1}\right\}\cdot \eta^2(1+\|\bm{\beta}\|_2^2) . $
In the sequel, we will verify the achievability of the equality in $(d)$ with two cases: $\bm{\beta} \in B_1$ and $\bm{\beta} \in B_1^c$.

\textbf{Case 1: $\bm{\beta} \in B_1$}:
For $\bm{\beta} \in B_1$, by taking $\tilde{\bm{x}}_0=\eta \frac{\bm{b}}{\|\bm{b}\|_2}$, we have
\begin{eqnarray}
	&&\hspace{-4mm}\frac{1}{m+1}(\|y_0-\bm{x}_0^T \bm{\beta}\|_2^2+\|\bm{y_1}-\bm{X}_1\bm{\beta}\|_2^2) \nonumber\\
	&& \hspace{10mm}-\frac{1}{n-m}\|\bm{y_2}-\bm{X}_2\bm{\beta}\|_2^2 \nonumber\\
		&&\hspace{-9mm}=\frac{1}{m+1}\left[\eta^2(1+\|\bm{\beta}\|_2^2)+\|\bm{y_1}-\bm{X}_1\bm{\beta}\|_2^2\right]\nonumber\\
	&& -\frac{1}{n-m}\|\bm{y_2}-\bm{X}_2\bm{\beta}\|_2^2 \nonumber\\
	&&\hspace{-9mm}\overset{(e)}{\geq}\max\left\{\frac{1}{2(m+1)}-\frac{1}{2\lambda(n+1)},0\right\} \eta^2(1+\|\bm{\beta}\|_2^2) \} , \label{eq:inner >0}
\end{eqnarray}
where $(e)$ is from the definition of set $B_1$. Then we have $L_1\overset{(f)}{=}g_1(\bm{\beta})$,
where $(f)$ follows from \eqref{eq:inner >0}. Therefore, for $ \bm{\beta} \in B_1$, we have $ h_1(\bm{\beta}) \leq g_1(\bm{\beta})$ and $L_1 \leq g_1(\bm{\beta})$, in which the equality can be achieved for $\tilde{\bm{x}}_0=\eta \frac{\bm{b}}{\|\bm{b}\|_2}$.

\textbf{Case 2: $\bm{\beta} \in B_1^c$}:
On the one hand, if $\frac{1}{m+1}-\frac{1}{\lambda(n+1)}\leq 0$, by taking $\tilde{\bm{x}}_0=\eta \frac{\bm{b}}{\|\bm{b}\|_2}$, we have
\begin{eqnarray}
	&&\frac{1}{m+1}(\|y_0-\bm{x}_0^T \bm{\beta}\|_2^2+\|\bm{y_1}-\bm{X}_1\bm{\beta}\|_2^2)\nonumber\\
	&&\hspace{10mm}-\frac{1}{n-m}\|\bm{y_2}-\bm{X}_2\bm{\beta}\|_2^2 \nonumber\\
	&\overset{(g)}{<}& \max\left\{\frac{1}{2(m+1)}-\frac{1}{2\lambda(n+1)},0\right\} \eta^2(1+\|\bm{\beta}\|_2^2) \}\nonumber\\
	&\overset{(h)}{=}&0, \label{eq:inner<0 1}
\end{eqnarray}
where $(g)$ is from the definition of set $B_1$ and $(h)$ is because $\frac{1}{m+1}-\frac{1}{\lambda(n+1)}\leq 0$. Then we have 
\begin{eqnarray}
	&& \hspace{-7mm} L_1\overset{(j)}{=}  \frac{1}{n+1}\left[\eta^2(1+\|\bm{\beta}\|_2^2)+\|\bm{y_1}-\bm{X}_1\bm{\beta}\|_2^2\right. \nonumber\\
	&&\hspace{7mm}\left. +\|\bm{y_2}-\bm{X}_2\bm{\beta}\|_2^2\right]\nonumber\\
	&&-\lambda \left[ \frac{1}{m+1}\eta^2(1+\|\bm{\beta}\|_2^2)+\frac{1}{m+1}\|\bm{y_1}-\bm{X}_1\bm{\beta}\|_2^2\right. \nonumber\\
	&& \hspace{7mm}\left. -\frac{1}{n-m}\|\bm{y_2}-\bm{X}_2\bm{\beta}\|_2^2\right]
	 \overset{(k)}{=} h_1(\bm{\beta}), \nonumber
\end{eqnarray}
where $(j)$ is from \eqref{eq:inner<0 1} and 
$(k)$ is true because $\frac{1}{m+1}-\frac{1}{\lambda(n+1)}\leq 0$. Therefore, for $ \bm{\beta} \in B_1^c$ and $\frac{1}{m+1}-\frac{1}{\lambda(n+1)} \leq 0$, we have $ h_1(\bm{\beta}) \geq g_1(\bm{\beta})$ and $L_1 \leq h_1(\bm{\beta})$, in which the equality can be achieved for $\tilde{\bm{x}}_0=\eta \frac{\bm{b}}{\|\bm{b}\|_2}$. 

On the other hand, if $\frac{1}{m+1}-\frac{1}{\lambda(n+1)}> 0$, by taking $\tilde{\bm{x}}_0$ to be a vector such that $\tilde{\bm{x}}_0 \perp \bm{b}$, we have
\begin{eqnarray}
	&&\hspace{-2mm}\frac{1}{m+1}(\|y_0-\bm{x}_0^T \bm{\beta}\|_2^2+\|\bm{y_1}-\bm{X}_1\bm{\beta}\|_2^2)\nonumber\\
	&& \hspace{21mm}-\frac{1}{n-m}\|\bm{y_2}-\bm{X}_2\bm{\beta}\|_2^2 \nonumber\\
	&&\hspace{-5mm}=\frac{1}{m+1}\|\bm{y_1}-\bm{X}_1\bm{\beta}\|_2^2-\frac{1}{n-m}\|\bm{y_2}-\bm{X}_2\bm{\beta}\|_2^2 \nonumber\\
	&&\hspace{-5mm}\overset{(l)}{<} \max\{-\frac{1}{2(m+1)}-\frac{1}{2\lambda(n+1)},-\frac{1}{m+1}\} \eta^2(1+\|\bm{\beta}\|_2^2) \}\nonumber\\
	&&\hspace{-5mm}<0, \label{eq:inner<0 2}
\end{eqnarray}
where $(l)$ is from the definition of set $B_1$. Then we have 
\begin{eqnarray}
	L_1&\overset{(s)}{=}&\frac{1}{n+1}\left(\|\bm{y_1}-\bm{X}_1\bm{\beta}\|_2^2+\|\bm{y_2}-\bm{X}_2\bm{\beta}\|_2^2\right) \nonumber\\
	&& \hspace{-5mm} -\lambda \left[ \frac{1}{m+1}\|\bm{y_1}-\bm{X}_1\bm{\beta}\|_2^2-\frac{1}{n-m}\|\bm{y_2}-\bm{X}_2\bm{\beta}\|_2^2\right]\nonumber\\
	&\overset{(t)}{=}& h_1(\bm{\beta}), \nonumber
\end{eqnarray}
where $(s)$ is from \eqref{eq:inner<0 2} and $(t)$ is because  $\frac{1}{m+1}-\frac{1}{\lambda(n+1)}> 0$. Therefore, for $ \bm{\beta} \in B_1^c$ and $\frac{1}{m+1}-\frac{1}{\lambda(n+1)} > 0$, we have $ h_1(\bm{\beta}) \geq g_1(\bm{\beta})$ and $L_1 \leq h_1(\bm{\beta})$, in which the equality can be achieved when $\tilde{\bm{x}}_0 \perp \bm{b}$.

\section{Proof of Proposition~\ref{prop:case 2}} \label{app:case 2 analysis}
First, we summarize the process of finding $\bar{\bm{\beta}}$ as follows. \\
\noindent 1. Check whether $A=\{ \alpha_1: \bm{M_{g_1}}-\alpha_1 ( \bm{M_{g_1}}-\bm{M_{h_1}})\succeq 0\} \neq \emptyset$. If $A=\emptyset$, there does not exist a global minimizer in this case.

\noindent 2. By randomly selecting an $\alpha_1^* \in A_{g_1h_1} \coloneqq \{ \alpha: \bm{M_{g_1}}-\alpha ( \bm{M_{g_1}}-\bm{M_{h_1}})\succ 0\}$, we solve the optimization problem
\begin{eqnarray}
	\min_{\bm{\beta}}&& k(\bm{\beta})=g_1(\bm{\beta})-\alpha_1^* [ g_1(\bm{\beta})-h_1(\bm{\beta})] , \nonumber\\
	\text{s.t. } && C_1(\bm{\beta})=g_1(\bm{\beta})-h_1(\bm{\beta}) = 0, \label{eq:min k(beta)}
\end{eqnarray}
where $k(\bm{\beta})$ is positive-definite and the choice of $\alpha_1^*$ does not affect the solution to the problem. 

\noindent 3. For the solution to \eqref{eq:min k(beta)}, check whether $\alpha_1>0$, \eqref{eq:sec-order 2}, \eqref{eq:fir-order 2} and \eqref{eq:opt ineq constraint 2} are satisfied. 

Now we explore the details of steps 1, 2 and 3. 

In step 1, the assumption $\eta^2 \geq \eta_{\min}^2= \max\left\{\frac{(n+1)v_{X_1,p}}{m(m+1)},\frac{(n+1)v_{X_2,p}}{(n-m+1)(n-m)}\right\}$ will guarantee that $A$ is nonempty.
To be exact, we denote 
	$A_{g_1g_2}=\{ \alpha: \bm{M_{g_1}}-\alpha ( \bm{M_{g_1}}-\bm{M_{g_2}})\succ 0\}, A_{g_1h_2}=\{ \alpha: \bm{M_{g_1}}-\alpha ( \bm{M_{g_1}}-\bm{M_{h_2}})\succ 0\}, A_{h_1g_2}=\{ \alpha: \bm{M_{h_1}}-\alpha ( \bm{M_{h_1}}-\bm{M_{g_2}})\succ 0\}, A_{h_1h_2}=\{ \alpha: \bm{M_{h_1}}-\alpha ( \bm{M_{h_1}}-\bm{M_{h_2}})\succ 0\}, A_{g_2h_2}=\{ \alpha: \bm{M_{g_2}}-\alpha ( \bm{M_{g_2}}-\bm{M_{h_2}})\succ 0\}.$
	Then under the assumption that $\eta^2 \geq \eta_{\min}^2= \max\left\{\frac{(n+1)v_{X_1,p}}{m(m+1)},\frac{(n+1)v_{X_2,p}}{(n-m+1)(n-m)}\right\}$, we are able to derive that $A_{g_1h_1} \neq \emptyset, A_{g_1g_2} \neq \emptyset,A_{g_1h_2} \neq \emptyset,A_{h_1g_2} \neq \emptyset,A_{h_1h_2} \neq \emptyset,A_{g_2h_2} \neq \emptyset$. The detailed proof is omitted here. Particularly, in this case study, we have $ A_{g_1h_1} \subset A$ and $A \neq \emptyset$. 

In step 2, \eqref{eq:min k(beta)} is a strictly convex quadratic optimization problem with one quadratic equality constraint, which has been discussed in \cite{hmam2010quadratic}. Define the Lagrangian function of \eqref{eq:min k(beta)} as
\begin{eqnarray}
	\mathcal{L}(\bm{\beta},\gamma)&=&k(\bm{\beta})-\gamma C_1(\bm{\beta}) \nonumber\\
	&=& g_1(\bm{\beta})-(\alpha_1^*+\gamma) (g_1(\bm{\beta})-h_1(\bm{\beta})) \nonumber\\
	&=& (1-\alpha_1^*-\gamma)g_1(\bm{\beta})+(\alpha_1^*+\gamma) h_1(\bm{\beta}),
	\nonumber
\end{eqnarray}
where $\gamma$ is the Lagrangian multiplier. According to \cite{hmam2010quadratic}, the global minimizer $\check{\bm{\beta}}$ and the corresponding multiplier $\gamma^*$ of \eqref{eq:min k(beta)} satisfy first-order, second-order and the constraint conditions
\begin{eqnarray}
	\frac{\partial\mathcal{L}}{\partial \bm{\beta}}\vert _{\check{\bm{\beta}}}&=&(1-\alpha_1^*-\gamma)\frac{\partial g_1(\bm{\beta})}{\partial \bm{\beta}} \vert_{\check{{\bm{\beta}}}}+(\alpha_1^*+\gamma) \frac{\partial h_1(\bm{\beta})}{\partial \bm{\beta}}\vert_{\check{{\bm{\beta}}}} \nonumber\\
	&=&\bm{0},\label{eq:first order}\\
	\frac{\partial^2\mathcal{L}}{\partial \bm{\beta}^2}&=&2\left[(1-\alpha_1^*-\gamma)\bm{M}_{g_1}+(\alpha_1^*+\gamma^*) \bm{M}_{h_1}\right]\succeq \bm{0}, \nonumber\\
	C_1( \check{\bm{\beta}})&=&0. \label{eq:constrain con}
\end{eqnarray}
From \eqref{eq:first order}, we have
\begin{eqnarray}
	\check{{\bm{\beta}}}&=&\left[(1-\alpha_1^*-\gamma^*)\bm{M}_{g_1}+(\alpha_1^*+\gamma^*) \bm{M}_{h_1}\right]^{-1} \nonumber\\
	&&\cdot \left[(1-\alpha_1^*-\gamma^*)\bm{E}_{g_1}-(\alpha_1^*+\gamma^*) \bm{E}_{h_1}\right]. \label{eq:beta check form}
\end{eqnarray}
Substituting \eqref{eq:beta check form} into \eqref{eq:constrain con}, we derive an equation for $\gamma$, $K(\gamma)=C_1( \check{\bm{\beta}})=0$, whose root is $\gamma^*$. By plugging $\gamma=\gamma^*$ back into \eqref{eq:beta check form}, the exact solution for $ \check{\bm{\beta}}$ is obtained.

For step 3, if $C_2(\check{\bm{\beta}}) \geq 0, C_3(\check{\bm{\beta}}) \geq 0$, then we have that: 
	\begin{enumerate}
		\item if $\alpha_1^*+\gamma^* >0$, $\bm{\beta}=\check{\bm{\beta}}$ is a global minimizer satisfying \eqref{eq:sec-order 2}, \eqref{eq:fir-order 2}, \eqref{eq:opt ineq constraint 2} with $\alpha_1=\alpha_1^*+\gamma^*$; 
		\item if $\alpha_1^*+\gamma^* =0$, $\bm{\beta}=\check{\bm{\beta}}$ is a global minimizer in \textbf{Case 1} that satisfies \eqref{eq:sec-order 1}, \eqref{eq:fir-order 1}, \eqref{eq:opt constraint 1}; 
		\item  if $\alpha_1^*+\gamma^* <0$, $\bm{\beta}=\check{\bm{\beta}}$ satisfies global optimality conditions for the minimization of $h_1({\bm{\beta}})$ with multipliers $\alpha_1'=1-\alpha_1^*-\gamma^*, \alpha_2'=\alpha_3'=0$.
	\end{enumerate}

\section{Proof of Proposition~\ref{prop:case 3}} \label{app:case 3 analysis}
First, we summarize the process of finding $\hat{\bm{\beta}}$ as follows. \\  
\noindent 1. Check $AA=\{ (\alpha_1,\alpha_2): \bm{M_{g_1}}-\alpha_1 ( \bm{M_{g_1}}-\bm{M_{h_1}})-\alpha_2 ( \bm{M_{g_1}}-\bm{M_{g_2}})\succeq 0\} \neq \emptyset$. Under the assumption made in Proposition~\ref{prop:case 2} that $\eta^2 \geq \eta_{\min}^2= \max\left\{\frac{(n+1)v_{X_1,p}}{m(m+1)},\frac{(n+1)v_{X_2,p}}{(n-m+1)(n-m)}\right\}$, we have $A_{g_1h_1}=\{ \alpha: \bm{M_{g_1}}-\alpha ( \bm{M_{g_1}}-\bm{M_{h_1}})\succ 0\} \neq \emptyset$ and $A_{g_1g_2}=\{ \alpha: \bm{M_{g_1}}-\alpha ( \bm{M_{g_1}}-\bm{M_{g_2}})\succ 0\} \neq \emptyset$, which implies $AA \neq \emptyset$. 

\noindent 2. Solve the optimization problem
\begin{eqnarray}
	\min_{\bm{\beta}}&& k(\bm{\beta})=g_1(\bm{\beta})-\alpha_1^* [ g_1(\bm{\beta})-h_1(\bm{\beta})], \nonumber\\
	\text{s.t. } && C_1(\bm{\beta}) = C_2(\bm{\beta}) = 0. \label{eq:min k(beta) with two constraints}
\end{eqnarray}

\noindent 3. For the solution to \eqref{eq:min k(beta) with two constraints}, check whether $\alpha_1>0, \alpha_2>0$, and \eqref{eq:opt ineq constraint double} are satisfied. 

We now provide more details of steps 2 and 3. In step 2, define the Lagrangian function of \eqref{eq:min k(beta) with two constraints} as
\begin{eqnarray}
	&&\mathcal{L}(\bm{\beta},\gamma_i)=k(\bm{\beta})-\gamma_1 C_1(\bm{\beta})-\gamma_2 C_1(\bm{\beta}) \nonumber\\
	&& \hspace{-5mm}= (1-\alpha_1^*-\gamma_1-\gamma_2)g_1(\bm{\beta})+(\alpha_1^*+\gamma_1) h_1(\bm{\beta})+\gamma_2 h_1(\bm{\beta}).
	\nonumber
\end{eqnarray}
Then the derived optimal solution $\check{\bm{\beta}}$ and the corresponding Lagrangian multipliers $\gamma_1^*,\gamma_2^*$ satisfy first-order, second-order and the constraint conditions
\begin{eqnarray}
	\frac{\partial\mathcal{L}}{\partial \bm{\beta}}\vert _{\check{\bm{\beta}}}&=&(1-\alpha_1^*-\gamma_1^*-\gamma_2^*)\frac{\partial g_1(\bm{\beta})}{\partial \bm{\beta}} \vert_{\check{{\bm{\beta}}}}\nonumber\\
	&&\hspace{-3mm}+(\alpha_1^*+\gamma_1^*) \frac{\partial h_1(\bm{\beta})}{\partial \bm{\beta}}\vert_{\check{{\bm{\beta}}}}+\gamma_2^* \frac{\partial g_2(\bm{\beta})}{\partial \bm{\beta}}\vert_{\check{{\bm{\beta}}}} =\bm{0},\label{eq:first order double}\\
	\frac{\partial^2\mathcal{L}}{\partial \bm{\beta}^2}&=&2[(1-\alpha_1^*-\gamma_1^*-\gamma_2^*)\bm{M}_{g_1}\nonumber\\
	&&+(\alpha_1^*+\gamma^*_1) \bm{M}_{h_1}+\gamma_2^* \bm{M}_{g_2}]\succeq \bm{0}, \label{eq:second order double} \\
	&& \hspace{-16mm}C_1( \check{\bm{\beta}})=0, C_2( \check{\bm{\beta}})=0. \label{eq:constrain con double}
\end{eqnarray}
From \eqref{eq:first order double}, we have
\begin{eqnarray}
	\bm{0}&=&\left[(1-\alpha_1^*-\gamma_1^*-\gamma_2^*)\bm{M}_{g_1} +(\alpha_1^*+\gamma_1^*)\bm{M}_{h_1} \right. \nonumber\\
	&& \left. +\gamma_2^*\bm{M}_{g_2}\right]\check{{\bm{\beta}}}-(1-\alpha_1^*-\gamma_1^*-\gamma_2^*)\bm{E}_{g_1}\nonumber\\
	&&-(\alpha_1^*+\gamma_1^*)\bm{E}_{h_1}-\gamma_2^*\bm{E}_{g_2}, \nonumber
\end{eqnarray}
where $\bm{E}_{g_2}=C_{g_2}\bm{X}_1^T\bm{y_1} +D_{g_2}\bm{X}_2^T\bm{y_2}$. Then we have 
\begin{eqnarray}
	&& \hspace{-7mm}\check{{\bm{\beta}}}=\left[(1-\alpha_1^*-\gamma_1^*-\gamma_2^*)\bm{M}_{g_1} +(\alpha_1^*+\gamma_1^*)\bm{M}_{h_1}+\gamma_2^*\bm{M}_{g_2}\right]^{-1}\nonumber\\
	&&\hspace{-2mm}\cdot\left[(1-\alpha_1^*-\gamma_1^*-\gamma_2^*)\bm{E}_{g_1}+(\alpha_1^*+\gamma_1^*)\bm{E}_{h_1}+\gamma_2^*\bm{E}_{g_2}\right]. \label{eq:beta check form double} 
\end{eqnarray}
Plugging \eqref{eq:beta check form double} into \eqref{eq:constrain con double}, we have
\begin{eqnarray}
	K_1(\gamma_1,\gamma_2)=C_1( \check{\bm{\beta}})=0, K_2(\gamma_1,\gamma_2)=C_2( \check{\bm{\beta}})=0, \nonumber
\end{eqnarray}
with solution $(\gamma_1^*,\gamma_2^*)$. By substituting $\gamma_1=\gamma_1^*,\gamma_2=\gamma_2^*$ into \eqref{eq:beta check form double}, we obtain the solution for $ \check{\bm{\beta}}$. 

For step 3, the verification process is given as follows. 
\begin{enumerate}
	\item If $\alpha_1^*+\gamma_1^*>0$ and $\gamma_2^*>0$, \eqref{eq:sec-order double}, \eqref{eq:fir-order double}, \eqref{eq:opt constraint double} are satisfied for $\alpha_1=\alpha_1^*+\gamma_1^*, \alpha_2=\gamma_2^*$ and $\bm{\beta}=\check{\bm{\beta}}$ based on \eqref{eq:first order double}, \eqref{eq:second order double}, \eqref{eq:constrain con double}. If we further have $C_3(\check{\bm{\beta}}) \geq 0$, then $ \check{\bm{\beta}}$ is a global minimizer of \eqref{eq:minimization g1}.
	\item If $\alpha_1^*+\gamma_1^*<0$, we could consider the minimization of $h_1(\bm{\beta})$.
	\item If $\gamma_2^*<0$, we consider the minimization of $g_2(\bm{\beta})$. 
\end{enumerate}

\section{Proof of Lemma~\ref{lm:max g D_g positive}} \label{app:max g D_g positive}
Note that $\bm{c}_1$ and $\bm{c}_2$ are independent without considering the optimization on $\eta_{c_1}$. In particular, the first term in $g(\bm{\beta},\hat{\bm{{X}}})$ only involves $\bm{c}_1$ and the second term in $g(\bm{\beta},\hat{\bm{{X}}})$ only involves $\bm{c}_2$. Thus, we firstly focus on the first term in $g(\bm{\beta},\hat{\bm{{X}}})$ and solve the maximization with respect to $\bm{c}_1$. 
\begin{eqnarray}
	&& \max_{\eta_{c_1}}\max_{\|\bm{d}\|_2\leq 1}\max_{\|\bm{c}_1\|_2 = \eta_{c_1}} \|\bm{y}_1-\bm{{X}}_1\bm{\beta}-\bm{c}_1\bm{d}^T\bm{\beta}\|_2^2 \nonumber\\
	&=&\max_{\eta_{c_1}}\max_{\|\bm{d}\|_2\leq 1}\max_{\|\bm{c}_1\|_2 = \eta_{c_1}} (\bm{d}^T\bm{\beta})^2 \|\bm{e}_1\|_2^2 \nonumber\\
	&=&\max_{\eta_{c_1}}\max_{\|\bm{d}\|_2\leq 1}(\bm{d}^T\bm{\beta})^2 \max_{\|\bm{c}_1\|_2 = \eta_{c_1}}  \|\bm{e}_1\|_2^2, \nonumber
\end{eqnarray}
in which $\bm{e}_1=\bm{f}_1-\bm{c}_1$ with $\bm{f}_1=\frac{1}{\bm{d}^T\bm{\beta}}(\bm{y}_1-\bm{X}_1\bm{\beta})$. For the maximization problem on $\bm{c}_1$, we have
	\begin{eqnarray}
		&& \max_{\bm{c}_1}  \|\bm{e}_1\|_2^2, ~~\text{s.t. }\|\bm{c}_1\|_2 = \eta_{c_1}, \nonumber\\
		&\Longleftrightarrow& \min_{\bm{e}_1}  -\|\bm{e}_1\|_2^2, ~~\text{s.t. }\|\bm{f}_1-\bm{e}_1\|_2^2 = \eta_{c_1}^2. \label{eq:min on e1}
	\end{eqnarray}
	Although \eqref{eq:min on e1} is not a convex optimization problem, we can first investigate its KKT necessary conditions. The Lagrangian function of \eqref{eq:min on e1} is $$\mathcal{L}(\bm{e}_1,\gamma_{\bm{e}_1})=-\|\bm{e}_1\|_2^2+\gamma_{\bm{e}_1}(\|\bm{f}_1-\bm{e}_1\|_2^2 -\eta_{c_1}^2),$$ 
	where $\gamma_{\bm{e}_1}$ is the Lagrangian multiplier. According to the KKT conditions, we have
	\begin{eqnarray}
		&&\partial	\mathcal{L}(\bm{e}_1,\gamma_{\bm{e}_1})=-2\bm{e}_1^T-2\gamma_{\bm{e}_1}(\bm{f}_1-\bm{e}_1)^T=\bm{0}, \nonumber\\
		&&\|\bm{f}_1-\bm{e}_1\|_2^2 =\eta_{c_1}^2, \nonumber
	\end{eqnarray}
	from which we can derive that the solution to \eqref{eq:min on e1} is $\bm{e}_1^*=\bm{f}_1+\frac{\eta_{c_1}}{\|\bm{f}_1\|_2}\bm{f}_1$, and the maximum value is $$\max_{\|\bm{c}_1\|_2 = \eta_{c_1}}  \|\bm{e}_1\|_2^2=\|\bm{e}_1^*\|_2^2=\left(1+\frac{\eta_{c_1}}{\|\bm{f}_1\|_2}\right)^2 \|\bm{f}_1\|_2^2.$$

Then we focus on the second term in $g(\bm{\beta},\hat{\bm{{X}}})$, solve the maximization on $\eta_{c_2}$, and derive the formulation for $g_{m_1}(\eta_{c_1},\bm{\beta},\bm{d})$.

\section{Proof of Proposition~\ref{prop:max g_a}} \label{app:max g_a}
We observe that $g_{m_1}(\eta_{c_1},\bm{\beta},\bm{d})$ is a quadratic function with respect to $\bm{d}^T\bm{\beta}$, i.e.
\begin{eqnarray}
	g_{m_1}(\eta_{c_1},\bm{\beta},\bm{d})=A(\bm{d}^T\bm{\beta})^2+B(\bm{d}^T\bm{\beta})+C, \label{eq:g_m form}
\end{eqnarray}
in which $A,B,C$ are three coefficients. In particular, we have
\begin{eqnarray}
	A&=&\left(C_g-D_g\right)\eta_{c_1}^2+D_g\eta_c^2, \label{eq:A}\\
	B&=&2\left[C_g\eta_{c_1}\|\bm{y}_1-\bm{X}_1\bm{\beta}\|_2+D_g\eta_{c_2}\|\bm{y}_2-\bm{X}_2\bm{\beta}\|_2\right]\geq0,\nonumber\\
	C&=&C_g\|\bm{y}_1-\bm{X}_1\bm{\beta}\|_2^2+D_g\|\bm{y}_2-\bm{X}_2\bm{\beta}\|_2^2. \label{eq:C}
\end{eqnarray} 
Since $A>0$, $-\frac{B}{2A}\leq0$ and $\bm{d}^T\bm{\beta} \in [-\|\bm{\beta}\|_2,\|\bm{\beta}\|_2]$, we can conclude that the maxima of $g_{m_1}(\bm{d} | \eta_{c_1},\bm{\beta})$ is attained when $\bm{d}^T\bm{\beta}=\|\bm{\beta}\|_2$ 
and the maximum value is 
\begin{eqnarray}
	&&\max_{\|\bm{d}\|_2 \leq 1} g_{m_1}(\eta_{c_1},\bm{\beta},\bm{d})\nonumber\\
	&=&C_g(\|\bm{y}_1-\bm{X}_1\bm{\beta}\|_2+\eta_{c_1}\|\bm{\beta}\|_2)^2\nonumber\\
	&&\hspace{-3mm}+D_g(\|\bm{y}_2-\bm{X}_2\bm{\beta}\|_2+\sqrt{\eta_c^2-\eta_{c_1}^2}\|\bm{\beta}\|_2)^2, \nonumber
\end{eqnarray}
which provides the form of $g_a$.

\section{Proof of Lemma~\ref{lm:max g D_g negative}} \label{app:max g D_g negative}
In this case, the analysis for the first term in $g(\bm{\beta},\hat{\bm{{X}}})$ remains the same. However, for the second term, we have  
\begin{eqnarray}
	&& \min_{\eta_{c_2}}\min_{\|\bm{d}\|_2\leq\frac{\eta}{\eta_{c_2}}}\min_{\|\bm{c}_2\|_2 \leq \eta_{c_2}} \|\bm{y}_2-\bm{{X}}_2\bm{\beta}-\bm{c}_2\bm{d}^T\bm{\beta}\|_2^2 \nonumber\\
	&=&\min_{\eta_{c_2}}\min_{\|\bm{d}\|_2\leq\frac{\eta}{\eta_{c_2}}}(\bm{d}^T\bm{\beta})^2  \min_{\|\bm{f}_2-\bm{e}_2\|_2 \leq \eta_{c_2}}  \|\bm{e}_2\|_2^2, \nonumber
\end{eqnarray}
where $\eta_{c_2}=\sqrt{\eta_c^2-\eta_{{c}_1}^2}$, $\bm{f}_2=\frac{1}{\bm{d}^T\bm{\beta}}(\bm{y}_2-\bm{X}_2\bm{\beta})$ and $\bm{e}_2=\bm{f}_2-\bm{c}_2$. Thus, the minimization on $\bm{e}_2$ is a convex problem. By exploring the KKT conditions of the minimization problem, we are able to find the optimal solution. Particularly, the Lagrangian function of the minimization problem on $\bm{e}_2$ is
\begin{eqnarray}
	\mathcal{L}(\bm{e}_2,\gamma_{\bm{e}_2})=\|\bm{e}_2\|_2^2+\gamma_{\bm{e}_2}(\|\bm{f}_2-\bm{e}_2\|_2^2 -\eta_{c_2}^2), \nonumber
\end{eqnarray}
in which $ \gamma_{\bm{e}_2}$ is the Lagrangian multiplier. By exploring the KKT conditions, we have 
\begin{eqnarray}
	&&\nabla	\mathcal{L}(\bm{e}_2,\gamma_{\bm{e}_2})=2\bm{e}_2^T-2\gamma_{\bm{e}_2}(\bm{f}_2-\bm{e}_2)^T=0, \label{eq:e_2 lagrangian gradient}\\
	&&\|\bm{f}_2-\bm{e}_2\|_2^2 \leq \eta_{c_2}^2, \nonumber\\
	&& \gamma_{\bm{e}_2}(\|\bm{f}_2-\bm{e}_2\|_2^2 -\eta_c^2)=0,\label{eq:complementary condition}\\
	&&  \gamma_{\bm{e}_2} \geq 0. \nonumber
\end{eqnarray}
By inspecting the complementary slackness condition \eqref{eq:complementary condition}, we consider two cases based on the value of $\gamma_{\bm{e}_2}$.

\textbf{Case 1:} $\gamma_{\bm{e}_2}=0$. In this case, we have $\bm{e}_2=\bm{0}$ according to \eqref{eq:e_2 lagrangian gradient}, which can be true when $\|\bm{f}_2\|_2 \leq \eta_{c_2}$. Moreover, note that
\begin{eqnarray}
	\|\bm{f}_2\|_2 \leq \eta_{c_2} \Longleftrightarrow& \|\bm{y}_2-\bm{X}_2\bm{\beta}\|_2 \leq \bm{d}^T \bm{\beta}\eta_{c_2} \nonumber\\
	&\hspace{-10mm}\overset{(a)}{\leq} \frac{\eta}{\eta_{c_2}}\|\bm{\beta}\|_2\eta_{c_2}=\eta \|\bm{\beta}\|_2,\nonumber
\end{eqnarray} 
where the equality in (a) is achieved if $\bm{d}=\frac{\eta}{\eta_{c_2} \|\bm{\beta}\|_2}\bm{\beta}$. Thus, as long as $\|\bm{y}_2-\bm{X}_2\bm{\beta}\|_2 \leq \eta \|\bm{\beta}\|_2$, the minimum value of $\|\bm{e}_2\|_2^2$ is 0.

\textbf{Case 2:} $\gamma_{\bm{e}_2}>0$. If there is no feasible solution in \textbf{Case 1}, we can conclude that $\|\bm{f}_2\|_2> \eta_{c_2}$. Moreover, by \eqref{eq:e_2 lagrangian gradient} and \eqref{eq:complementary condition}, we have $\bm{e}^*_2=\frac{\gamma^*_{\bm{e}_2}\bm{f}_2}{\gamma^*_{\bm{e}_2}+1},~ \eta_{c_2}=\|\bm{f}_2-\bm{e}^*_2\|_2 =\frac{1}{\gamma_{\bm{e}^*_2}+1}\|\bm{f}_2\|_2$, 
which implies 
$\gamma^*_{\bm{e}_2}=\frac{\|\bm{f}_2\|_2}{\eta_{c_2}}-1,
	~\bm{e}^*_2=\bm{f}_2-\frac{\eta_{c_2}}{\|\bm{f}_2\|_2}\bm{f}_2.$
Then we have
\begin{eqnarray}
	\min_{\|\bm{f}_2-\bm{e}_2\|_2 \leq \eta_{c_2}}  \|\bm{e}_2\|_2^2= \|\bm{e}^*_2\|_2^2=\left(1-\frac{\eta_{c_2}}{\|\bm{f}_2\|_2}\right)^2 \|\bm{f}_2\|_2^2. \nonumber
\end{eqnarray}
By combining these two cases, Lemma~\ref{lm:max g D_g negative} is proved. 

\section{Proof of Proposition~\ref{prop:max g_b}} \label{app:max g_b}
Now we solve the maximization problem on $\bm{d}$. Firstly, consider the case when $\| \bm{y}_2-\bm{X}_2 \bm{\beta}\|_2 \leq \eta \|\bm{\beta}\|_2$. In this case, we notice that as long as $\eta_{{c}_1} \neq 0$, $g_{m_2}(\eta_{c_1},\bm{\beta},\bm{d})$ is a quadratic function for $\bm{d}^T\bm{\beta}$ with $A=C_g\eta_{c_1}^2>0$, $B=2C_g\eta_{c_1}\|\bm{y}_1-\bm{X}_1\bm{\beta}\|_2\geq 0$ and $-\frac{B}{2A}\leq0$. Thus, the maxima is attained when $\bm{d}^T\bm{\beta}=\|\bm{\beta}\|_2$ and the maximum value of $g(\bm{\beta},\bm{\hat{X}})$ is 
\begin{eqnarray}
	&g_{b_1}(\eta_{c_1},\bm{\beta})=C_g(\|\bm{y}_1-\bm{X}_1\bm{\beta}\|_2+\eta_{c_1}\|\bm{\beta}\|_2)^2.\nonumber
\end{eqnarray}
For $\eta_{c_1} = 0$, the attacker only changes the feature matrix of the second group and the maximum value of $g(\bm{\beta},\bm{\hat{X}})$ can also be derived as $g_{b_1}(\eta_{c_1},\bm{\beta})$.

Secondly, consider the case when $\| \bm{y}_2-\bm{X}_2 \bm{\beta}\|_2 > \eta \|\bm{\beta}\|_2$. In this case, $g_{m_2}(\eta_{c_1},\bm{\beta},\bm{d})$ can also be written in the form of \eqref{eq:g_m form} with coefficients $A, B, C$. In particular, $A$ and $C$ are defined the same as \eqref{eq:A} and \eqref{eq:C}, and $B$ is defined as $B=2C_g\eta_{c_1}\|\bm{y}_1-\bm{X}_1\bm{\beta}\|_2-2D_g\eta_{c_2}\|\bm{y}_2-\bm{X}_2\bm{\beta}\|_2\geq 0$. Since the coefficient of the quadratic term $A$ can be positive, negative or zero, the maxima of $g_{m_2}$ varies. By investigating into these three different cases, we have that when $\| \bm{y}_2-\bm{X}_2 \bm{\beta}\|_2 > \eta \|\bm{\beta}\|_2$, the maximum value of $g(\bm{\beta},\bm{\hat{X}})$ is $g_{b_2}(\eta_{c_1},\bm{\beta})$. 

If $A>0$, we have $-\frac{B}{2A}\leq0$ and the maxima is attained when $\bm{d}^T\bm{\beta}=\|\bm{\beta}\|_2$ with the maximum value to be
$\max_{\|\bm{d}\|_2 \leq 1} g_{m_2}(\eta_{c_1},\bm{\beta},\bm{d})=C_g(\|\bm{y}_1-\bm{X}_1\bm{\beta}\|_2+\eta_{c_1}\|\bm{\beta}\|_2)^2+D_g(\|\bm{y}_2-\bm{X}_2\bm{\beta}\|_2-\eta_{c_2}\|\bm{\beta}\|_2)^2,$
which implies that
\begin{eqnarray}
	&&\max_{\|\bm{c}\bm{d}^T \|_F \leq \eta} g(\bm{\beta},\bm{\hat{X}})\nonumber\\
	&=&\max_{0 < \eta_{c}\leq \eta}\max_{0 \leq \eta_{c_1}\leq \eta_c}\max_{\|\bm{d}\|_2 \leq 1} g_{m_2}(\eta_{c_1},\bm{\beta},\bm{d}) \nonumber\\
	&\overset{(a)}{=}&\max_{0 \leq \eta_{c_1}\leq \eta}\left[C_g(\|\bm{y}_1-\bm{X}_1\bm{\beta}\|_2+\eta_{c_1}\|\bm{\beta}\|_2)^2 \right.\nonumber\\
	&&\hspace{-7mm}\left. + D_g(\|\bm{y}_2-\bm{X}_2\bm{\beta}\|_2-\max_{0 < \eta_{c}\leq \eta}\eta_{c_2}\|\bm{\beta}\|_2)^2\right]\nonumber\\
	&=& \max_{0 \leq \eta_{c_1}\leq \eta}g_{b_2}(\eta_{c_1},\bm{\beta}), \nonumber 
\end{eqnarray}
where (a) follows from the fact that $D_g<0$ and $\| \bm{y}_2-\bm{X}_2 \bm{\beta}\|_2 > \eta \|\bm{\beta}\|_2 \geq \eta_{c_2} \|\bm{\beta}\|_2$. 

If $A=0$, from the expression of $A$, we have $\eta_{c_1}^2=\frac{-\frac{1}{n}+\frac{\lambda}{n-m}}{\frac{\lambda}{m}+\frac{\lambda}{n-m}} \eta_c^2$, which is feasible as $\frac{-\frac{1}{n}+\frac{\lambda}{n-m}}{\frac{\lambda}{m}+\frac{\lambda}{n-m}} \in (0,1)$. Then since $B\geq0$, $g_{m_2}$ is a linearly non-decreasing function in $\bm{d}^T\bm{\beta}$ and the maxima is attained when $\bm{d}^T\bm{\beta}=\|\bm{\beta}\|_2$ with the maximum value to be the same as $g_{b_2}(\eta_{c_1},\bm{\beta})$. 

Otherwise, if $A<0$, $g_{m_2}$ is a concave quadratic function in $\bm{d}^T\bm{\beta}$ with $-\frac{B}{2A}>\frac{\left(\frac{\lambda}{n-m}-\frac{1}{n}\right)\eta_{c_2}\|\bm{y}_2-\bm{X}_2\bm{\beta}\|_2}{-\left(\frac{\lambda}{n-m}-\frac{1}{n}\right)\eta_{c_1}^2+\left(\frac{\lambda}{n-m}-\frac{1}{n}\right)\eta_c^2}\overset{(g)}{>}\frac{\eta_{c_2}\eta \|\bm{\beta}\|_2}{\eta_{c_2}^2} \geq \|\bm{\beta}\|_2, $
in which (g) is from the fact that $\| \bm{y}_2-\bm{X}_2 \bm{\beta}\|_2 > \eta \|\bm{\beta}\|_2$. Thus, the maxima is attained when $\bm{d}^T\bm{\beta}=\|\bm{\beta}\|_2$ and the maximum value is also $g_{b_2}(\eta_{c_1},\bm{\beta})$. 

\section{Proof of Lemma~\ref{lm:weakly convex concave}} \label{app:weakly convex concave}
Since the forms of $g_a, g_{b_1},g_{b_2},h_a,h_{b_1},h_{b_2}$ are similar, we only show the weakly-convex-weakly-concave property of $g_{a}$. For $\eta_{{c}_1}$, we have
\begin{eqnarray}
	&&\hspace{-5mm}\frac{\partial^2 g_{a}(\eta_{{c}_1},\bm{\beta}) }{\partial \eta_{c_1}^2}\nonumber\\
	&=&2\left(\frac{\lambda}{m}+\frac{\lambda}{n-m}\right)\|\bm{\beta}\|_2^2 -2D_g\frac{\eta^2}{\eta_{c_2}^3}\|\bm{\beta}\|_2\|\bm{y}_2-\bm{X}_2\bm{\beta}\|_2. \nonumber
\end{eqnarray}
Since $D_g \geq 0$, as long as $\|\bm{\beta}\|_2$ is bounded,
there always exist a constant $\rho_1 < \infty$ such that $\frac{\partial^2 g_{a}(\eta_{{c}_1},\bm{\beta}) }{\partial \eta_{c_1}^2} \leq \rho_1$, which indicates that $g_a$ is weakly-concave in $\eta_{c_1}$. 

For $\bm{\beta}$, we have
\begin{eqnarray}
	\frac{\partial^2 g_{a}(\eta_{{c}_1},\bm{\beta})}{\partial \bm{\beta}^2} 
	\geq 2C_g\left[\eta_{c_1}\left(\eta_{c_1}- 2\frac{\mathrm{Tr}(\bm{X}_1^T\bm{X}_1)}{\|\bm{X}_1\|_F}\right) +\bm{X}_1^T\bm{X}_1\right] \nonumber\\
    \hspace{-5mm}+2D_g\left[\eta_{c_2}\left(\eta_{c_2}- 2\frac{\mathrm{Tr}(\bm{X}_2^T\bm{X}_2)}{\|\bm{X}_2\|_F}\right) +\bm{X}_2^T\bm{X}_2\right]. \nonumber
\end{eqnarray}
Since $\bm{X}_1$ and $\bm{X}_2$ are feature matrices with finite norm, there always exist $\rho_2 < \infty$ such that $\frac{\partial^2 g_{a}(\eta_{{c}_1},\bm{\beta}) }{\partial \bm{\beta}^2} \geq -\rho_2\bm{I}$, which indicates that $g_a$ is weakly-convex in $\bm{\beta}$. 

\section{Proof of Lemma~\ref{lm:unimodal}} \label{app:unimodal}
For $g_a$, we have
\begin{eqnarray}
	&&\frac{\partial g_{a}(\eta_{{c}_1},\bm{\beta}) }{\partial \eta_{c_1}}
	=2C_g\|\bm{\beta}\|_2\|\bm{y}_1-\bm{X}_1\bm{\beta}\|_2+\nonumber\\
	&&\hspace{-5mm}2\left(C_g-D_g\right)\eta_{c_1}\|\bm{\beta}\|_2^2 -2D_g\frac{\eta_{c_1}}{\eta_{c_2}}\|\bm{\beta}\|_2\|\bm{y}_2-\bm{X}_2\bm{\beta}\|_2 =0,\nonumber
\end{eqnarray}
which implies
\begin{eqnarray}
	&&\hspace{-5mm}\left(\eta_{c_1}\|\bm{\beta}\|_2+\frac{C_g}{C_g-D_g}\|\bm{y}_1-\bm{X}_1\bm{\beta}\|_2\right) \nonumber\\
	&&\cdot \left(\eta_{c_2}\|\bm{\beta}\|_2-\frac{D_g}{C_g-D_g}\|\bm{y}_2-\bm{X}_2\bm{\beta}\|_2\right) \nonumber\\
	&&\hspace{-10mm}=-\frac{C_gD_g}{\left(\frac{\lambda}{n-m}+\frac{\lambda}{m}\right)^2}\|\bm{y}_1-\bm{X}_1\bm{\beta}\|_2\|\bm{y}_2-\bm{X}_2\bm{\beta}\|_2.
	\label{eq:derivative}
\end{eqnarray}
From \eqref{eq:derivative}, we note that $\eta_{c_1}$ and $\eta_{c_2}$ are inversely proportional. Since we also have $\eta_{c_1}^2+\eta_{c_2}^2=\eta^2$, $\eta_{c_1}\geq 0$ and $\eta_{c_2} \geq 0$, there is a unique solution for \eqref{eq:derivative} (which can be seen geometrically), denoted as $\eta_{c_1}^*$. Moreover, we have
\begin{itemize}
	\item $\eta_{c_1} < \eta_{c_1}^*$, left hand side of \eqref{eq:derivative} is positive;
	\item $\eta_{c_1} > \eta_{c_1}^*$, left hand side of \eqref{eq:derivative} is negative.
\end{itemize}
Thus, $g_a$ is a unimodal function that increases first and then decreases. The results can be easily generalized to other sub-functions. 

\bibliographystyle{ieeetr}
\bibliography{ref}

\end{document}